\pdfoutput=1
\RequirePackage{ifpdf}
\ifpdf % We are running pdfTeX in pdf mode
\documentclass[pdftex]{sigma}
\else
\documentclass{sigma}
\fi

\usepackage{tikz}
\usepackage[all]{xy}

\numberwithin{equation}{section}

\newtheorem{Theorem}{Theorem}[section]
\newtheorem{Lemma}[Theorem]{Lemma}
\newtheorem{Proposition}[Theorem]{Proposition}
 { \theoremstyle{definition}
\newtheorem{Definition}[Theorem]{Definition}
\newtheorem{Remark}[Theorem]{Remark} }

\newcommand{\alg}[1]{\mathfrak{#1}}

\newcommand{\affine}[1]{\widehat{#1}}

\newcommand{\SLSA}[3]{\alg{#1} \left( #2 \middle\vert #3 \right)}
\newcommand{\AKMSA}[3]{\affine{\alg{#1}} \left( #2 \middle\vert #3 \right)}
\newcommand{\tgl}{\widetilde{\SLSA{gl}{1}{1}}}
\newcommand{\tgla}{\SLSA{\hat{\widetilde{gl}}}{1}{1}}
\newcommand{\tsl}{\widetilde{\SLSA{sl}{1}{1}}}
\newcommand{\gl}{{\SLSA{gl}{1}{1}}}

\begin{document}

\newcommand{\arXivNumber}{1411.1072}

\allowdisplaybreaks

\renewcommand{\PaperNumber}{067}

\FirstPageHeading

\ShortArticleName{Harmonic Analysis and Free Field Realization of the Takif\/f Supergroup of ${\rm GL}(1|1)$}

\ArticleName{Harmonic Analysis and Free Field Realization\\ of the Takif\/f Supergroup of $\boldsymbol{{\rm GL}(1|1)}$}

\Author{Andrei BABICHENKO~$^\dag$ and Thomas CREUTZIG~$^\ddag$}

\AuthorNameForHeading{A.~Babichenko and T.~Creutzig}

\Address{$^\dag$~Department of Mathematics, Weizmann Institut, Rehovot, 76100, Israel}
\EmailD{\href{mailto:babichenkoandrei@gmail.com}{babichenkoandrei@gmail.com}}

\Address{$^\ddag$~Department of Mathematical and Statistical Sciences, University of Alberta,\\
\hphantom{$^\ddag$}~Edmonton, Alberta T6G 2G1, Canada}
\EmailD{\href{mailto:creutzig@ualberta.ca}{creutzig@ualberta.ca}}

\ArticleDates{Received May 28, 2015, in f\/inal form August 01, 2015; Published online August 06, 2015}

\Abstract{Takif\/f superalgebras are a family of non semi-simple Lie superalgebras that are believed to give rise to a rich structure of indecomposable representations
of associated conformal f\/ield theories. We consider the Takif\/f superalgebra of~$\mathfrak{gl}(1\vert 1)$, especially we perform harmonic analysis for the corresponding supergroup.
We f\/ind that every simple module appears as submodule of an inf\/inite-dimensional indecomposable but reducible module.
We lift our results to two free f\/ield realizations for the corresponding conformal f\/ield theory and construct some modules.}


\Classification{17B67; 17B81; 22E46; 81R10; 81T40}

\section{Introduction}

Logarithmic conformal f\/ield theories carry this name as correlation functions sometimes have logarithmic singularities. They are playing an essential role in dif\/ferent
physical problems ranging from string theory,
especially on supergroup target spaces~\cite{ABT2, BT,CCMV,CMQSS,GQS, MQS, MQS1, QS, CQS2, RS,SS1} to dif\/ferent condensed matter and statistical mechanics problems~\cite{GJSV,GRS1,GRS,MR, PRZ,PR,ReS,ReS1,VJS}. Many of these statistical mechanics problems are described by logrithmic CFTs based on Lie superalgebras, as, e.g., supersymmetric disordered systems~\cite{GLL, Z}. For reviews on logarithmic CFT, see~\cite{CR3, F, Ga}.

The presence of logarithmic singularities is tightly connected to the non semi-simple action of the chiral algebra on some of its modules. In some models~\cite{ABT} of {\it non chiral} conformal theory related to string theory on ${\rm AdS}_3\times S^3$ non semi-simplicity appears on the level of the assumed symmetry, and not only the representation level. In~\cite{BR}, motivated by this fact, dif\/ferent {\it chiral} conformal f\/ield theories were considered, where non semi-simplicity appears on the level of symmetry algebra itself. These are conformal f\/ield theories based on Takif\/f (super)algebras, e.g., algebras that are non semi-simple extensions of simple Lie (super)algebras by its adjoint representation. Detailed algebraic structure of its representation theory was then investigated.

In the mathematics literature, this type of non-semisimple Lie superalgebras
were introduced by Takif\/f~\cite{T}, though not in the super setting, but as part of an investigation of invariant polynomial rings. These algebras have since been considered in a slightly generalised form under the names generalised Takif\/f algebras~\cite{G,W} in which a semisimple Lie algebra is tensored with a~polynomial ring in a nilpotent
formal variable $t$, and truncated current algebras~\cite{W1, W} in which one does the same to an af\/f\/ine Kac--Moody algebra. The algebras considered in~\cite{BR} correspond to taking
$t^2=0$, as in Takif\/f's original paper, and they were named Takif\/f superalgebras. These algebras were mainly considered algebraically from the representation theory point of view, in their chiral sector. On the other hand, it is well known that potential additional algebraic structures appear for Wess--Zumino--Novikov--Witten (WZNW) theories when representations are subjected to both chiral and anti chiral action of the group. In the present work, we initiate the study of the WZNW theory for the Takif\/f Lie supergroup~$\widetilde{{\rm GL}(1|1)}$.
In~\cite{SO2, SO1, Mo},  WZNW theories of non-reductive Lie groups have been studied and our notion of Takif\/f superalgebra is what the authors of~\cite{SO1} there call a double extensions.

Wess--Zumino--Novikov--Witten theories of non-compact Lie groups and Lie supergroups provide a rich source
of logarithmic conformal f\/ield theories. Usually, one considers such theories based on a semi-simple Lie (super)group or closely related supergroups as ${\rm GL}(n|n)$.\footnote{Even though $\SLSA{gl}{n}{n}$ is not semi-simple its representation theory is similar to the one of the simple Lie superalgebra $\SLSA{psl}{n}{n}$. However, already in the case of $\SLSA{gl}{1}{1}$ representations of the Takif\/f superalgebra have much more involved indecomposability structure than the non-Takif\/f superalgebra~\cite{BR}.}
Expe\-rience shows that in such theories most modules are actually completely reducible and only a few non-generic modules are indecomposable
but reducible. It is an interesting situation if already the Lie (super)algebra/Lie (super)group on which the conformal f\/ield theory is based is itself non semi-simple, as, e.g., for Takif\/f (super)groups.
In~\cite{BR}, the f\/irst author and David Ridout started to study such algebras from a conformal f\/ield theory perspective. Here, we'd like to continue their work with the aim of studying an example of a conformal f\/ield theory based on a Takif\/f superalgebra in detail. We believe that the case of~$\tgl$ (the Takif\/f superalgebra of the Lie superalgebra~$\gl$) is both treatable and instructive, so that we decided to f\/irst focus on this case. The chiral algebra of a WZNW model of a Lie group at level $k$ is the simple af\/f\/ine vertex operator algebra of its Lie algebra~$\mathfrak g$ at level~$k$. Frenkel and Zhu~\cite{FZ}  have proven that there is a one-to-one correspondence of simple representations of this vertex algebra and simple representations of a quotient of the universal envelopping algebra of the Lie algebra~$\mathfrak g$. If the vacuum Verma module of the af\/f\/ine vertex operator algebra is simple then there is actually a~one-to-one correspondence to the simple objects of the universal envelopping algebra. As in the case of $\gl$, the vacuum Verma module of the vertex algebra of $\tgl$ is simple for non-vanishing levels~\cite{BR}. So, we expect that representations of $\tgl$ can teach us quite a bit about the conformal f\/ield theory. Very natural modules of the Lie superalgebra are functions on the Lie supergroup which are studied using harmonic analysis (often called mini superspace analysis in the CFT literature) on the Lie supergroup.
This is actually a very general experience in WZNW theories based on Lie supergroups. The key strategy is to understand a~problem of the f\/inite-dimensional Lie superalgebra, to lift this understanding to the full conformal f\/ield theory, usually to a free f\/ield realization, and then to use the new understanding to derive inte\-res\-ting statements concerning the CFT.
This strategy has been initiated by Hubert Saleur, Volker Schomerus and Thomas Quella a decade ago. It has been used extensively by various people, including the second author. For example, the key starting ingredient of the bulk theories of $\text{GL}(1|1)$~\cite{SS1}, $\text{SU}(2|1)$~\cite{SS2}, $\text{PSU}(1,1|2)$~\cite{GQS} as well as the more general case~\cite{QS} has been the harmonic analysis, and then lifting the f\/indings to an appropriate free f\/ield realization.
These free f\/ield realizations in turn were very useful in computing correlation functions~\cite{SS1}, proving dualities to super Liouville f\/ield theories~\cite{CHR, HS}
and f\/inding relations to CFTs with $\mathcal N=(2, 2)$ world-sheet superconformal symmetry~\cite{CRo2}.
Similarly also boundary theories based on $\text{GL}(1|1)$~\cite{CQS, CS}  and $\text{OSP}(1|2)$~\cite{CH} have been studied by f\/irst performing an appropriate harmonic ana\-lysis.
The dif\/ference to the bulk theory is, that instead of studying the action of both left- and right-invariant vector f\/ields one investigates a twisted adjoint action. Boundary states are then lifts of distributions localized on twisted superconjugacy classes~\cite{C2, CQS}.

The by far best understood WZNW theory of a Lie supergroup is the one of $\text{GL}(1|1)$. It has been f\/irst studied by Rozansky and Saleur a while ago~\cite{RS}. Then as mentioned, Schomerus and Saleur, were able to understand spectrum and correlation functions using harmonic analysis and free f\/ield realization~\cite{SS1}.
As the next step boundary states were constructed~\cite{CQS}. Studying the twisted adjoint action of invariant vector f\/ields on functions on the supergroup allowed one to f\/ind the boundary states explicitely. Boundary and bulk-boundary correlation functions could also be computed using a similar free f\/ield realization as in the bulk case~\cite{CS}. The indecomposability structure of modules is not directly visible in the free f\/ield realization used in the just mentioned works. It turns out that there is another free f\/ield realization using symplectic fermions that makes this structure much clearer~\cite{CRo, LeC}. Using then the work of Kausch on symplectic fermions~\cite{K} most of the previous results could be obtained fairly directly~\cite{CRo}. This conformal f\/ield theory can also be studied from a more algebraic perspective, especially many extended algebras like ${\mathfrak{sl}}(2|1)$ at levels~$-1/2$ and~$1$, can be constructed and studied~\cite{AC, CR1, CR2}. In summary, the ${\rm GL}(1|1)$ WZNW theory has been thoroughly studied and has revealed a rich structure. We believe that a similar (but much more complex) story holds for its Takif\/f superalgebra. In~\cite{BR}, representations of both f\/inite-dimensional and af\/f\/ine Takif\/f superalgebra of $\gl$ have been studied and a Verlinde fusion ring has been computed. We now would like to proceed to the next step, and ask which representations appear in the full bulk CFT, and wether we can realize them using free f\/ields. Our main result is a thorough harmonic analysis of the Takif\/f supergroup. We f\/ind that functions split into three classes, called typical, semitypical and atypical. All of them appear in the harmonic analysis as submodule of inf\/inite-dimensional indecomposable but reducible modules.
The reason for this is a special element~$\tilde N$ of the Lie superalgebra, that already acts non-semi-simple in the adjoint representation.  It forces us to allow for a larger space of functions.
Nonetheless, in the typical case there is still a~decomposition into tensor products of modules for the action of left- and right-invariant vector f\/ields somehow similar as one is used from the Peter--Weyl theorem for compact Lie groups.
Both semitypical and especially atypical modules have a rather complicated behaviour under the left-right action. We visualize the structure of these modules in various pictures and exact sequences.
 Further, already the simple submodules of the typical modules posses Jordan cells of size three under the action of the Laplacian (as expected from~\cite{BR}). The harmonic analysis tells us nicely how to construct a free f\/ield realization including screening charges. As in the case of the ${\rm GL}(1|1)$ WZNW theory, typical modules can be constructed easily using this free f\/ield realization. Again, in analogy to the ${\rm GL}(1|1)$ WZNW theory we f\/ind a second free f\/ield realization using symplectic fermions which allows us to also construct semitypical and atypical modules. This puts us now in a position to study (in the near future) these modules further, to ask questions about correlation functions, operator product algebra and extended algebras.

This work is organized as follows:
Section~\ref{section2} is the main part of this paper. We f\/irst compute basic objects on the Takif\/f supergroup, namely invariant vector f\/ields, invariant measure and Laplacians. Then we compute the action of invariant vector f\/ields. The typical case can be performed rather directly, while the structure in the semitypical and atypical cases is very rich and f\/irst studying the action of various subalgebras turns out to structure the problem nicely.
Section~\ref{section3} then applies our f\/indings to construct a free f\/ield realization including screening charges. But we also add another free f\/ield realization using symplectic fermions.
Finally, we conclude with describing the expected future use of our f\/indings.

\section{Harmonic analysis}\label{section2}

We start from def\/inition of Takif\/f superalgebra $\widetilde{\text{GL}(1|1)}$, and brief\/ly describe its highest weight representations.
 The main part of this section is devoted to harmonic analysis for $\widetilde{\text{GL}(1|1)}$ supergroup. The procedure for this follows ideas established in various works on ${\rm GL}(1|1)$ (see mainly~\cite{SS1}) and goes as follows:
\begin{enumerate}\itemsep=0pt
\item[1)] consider a supergroup element $g$ according to a triangular decomposition;
\item[2)] compute left- and right-invariant vector f\/ields;
\item[3)] compute the invariant measure;
\item[4)] compute the Laplacians;
\item[5)] decompose functions with respect to the combined left-right action of invariant vector f\/ields.
\end{enumerate}

\subsection[Takiff superalgebra $gl(1|1)$ and its highest weight representations]{Takif\/f superalgebra $\boldsymbol{\tgl}$ and its highest weight representations}

Takif\/f Lie superalgebra $\tgl$ was introduced in~\cite{BR} as
an  extention of the superalgebra $\gl$ with Grassmann even
generator $N$, Grassmann odd generators $\psi ^{\pm }$ and central element $E$
\begin{gather*}
 [N,\psi^\pm]=\pm \psi^\pm ,\qquad \{\psi^+,\psi^-\}=E
\end{gather*}
by a set of their Takif\/f partners $\tilde{N}$, $\tilde{\psi }^{\pm }$, $\tilde{E}$ with commutation relations
\begin{gather*}
[N,\tilde{\psi}^\pm]=[\tilde{N},\psi^\pm]=\pm
\tilde{\psi}^\pm,\qquad \{\tilde{\psi}^+,\psi^-\}=\{\psi^+,
\tilde{\psi}^-\}=\tilde{E}
\end{gather*}
and the remaining (anti)commutators vanish.

As one can see the adjoint action of $N$, $E$, $\tilde{E}$ is diagonalised in
the chosen basis of generators, but the adjoint action of~$\widetilde{N}$
acts non semisimply on adjoint module. The obvious triangular decomposition
\begin{gather*}
\widetilde{\mathfrak{gl}(1|1)}=\text{span}\big\{\psi^{-},\tilde{\psi}^{-}\big\}\oplus
\text{span}\big\{N,E,\tilde{N},\tilde{E}\big\}\oplus \text{span}\big\{\psi^{+},
\widetilde{\psi}^{+}\big\}
\end{gather*}
gives rise to highest weight modules: the highest weight is def\/ined as an
eigenvector of $N$, $E$, $\tilde{N}$, $\tilde{E}$ with eigenvalues $n$, $e$, $\tilde{n}$, $\tilde{e}$ correspondingly, which is annihilated by the
action of raising genera\-tors~$\psi^{+}$,~$\tilde{\psi}^{+}$. The highest
weight Verma module is def\/ined as a free action of lowering genera\-tors~$\psi
^{-}$,~$\tilde{\psi}^{-}$ on the highest weight. Since~$\psi^{-}$, $\tilde{\psi}^{-}$ are Grassmann odd and anticommute, the Verma module
is four-dimensional. All the states of the Verma module with highest weight $|v\rangle $ are eigenstates of~$\tilde{N}$ except for $\psi
^{-}|v\rangle $ on which it acts non semisimply: $\tilde{N}\psi
^{-}|v\rangle =\tilde{n}\psi^{-}|v\rangle -\tilde{\psi}^{-}|v\rangle $.  As it was shown in~\cite{BR}, there are three
possibilities for irreducible quotients of Verma module. If $e=\tilde{e}=0$, the irreducible quotient is one-dimensional and was called \textit{%
atypical}, if $\tilde{e}=0$, but $e\neq 0$, it is two-dimensional and
was called \textit{semitypical}, and if~$\tilde{e}\neq 0$, the Verma
module is irreducible. In a similar way one can def\/ine lowest weight Verma
modules and their irreducible quotients.

There are two linearly independent quadratic Casimir operators in the
universal enveloping algebra of~$\tgl$ (modulo polynomials in
central elements) which can be chosen as
\begin{gather*}
Q_{1}=N\tilde{E}+\tilde{N}E+\psi^{-}\tilde{\psi}^{+}+%
\tilde{\psi}^{-}\psi^{+},\qquad Q_{2}=\tilde{N}\tilde{E}+\tilde{\psi}^{-}\tilde{\psi }^{+}.
\end{gather*}
They act on the highest weight Verma module as multiplication by $n
\tilde{e}+\tilde{n}e$, $\tilde{n}\tilde{e}$ respectively.

An important object for af\/f\/inization of Lie (super)algebras is a~non-degenerate symmetric bilinear form $\kappa $ on the algebra. Among
dif\/ferent possibilities, we choose the lifting of the standard
bilinear form $\kappa _{0}(X,Y)$ of $\mathfrak{gl}(1|1)$ (def\/ined as supertrace $\operatorname{str}(XY)
$ in the def\/ining representation of it) as follows
\begin{gather*}
\tilde{\kappa }(\tilde{X},Y)=\kappa_{0}(X,Y),\qquad \tilde{\kappa}(\tilde{X},\tilde{Y})=\tilde{\kappa}(X,Y)=0.
\end{gather*}

\subsection{Invariant vector f\/ields}

We choose a Lie supergroup parameterization according to above triangular decomposition
\begin{gather}\label{eq:triangular}
g = e^{i\theta_+\psi^++i\tilde\theta_+\tilde\psi^+}e^{ixE+iyN+i\tilde x\tilde E+i\tilde y\tilde N} e^{i\theta_-\psi^-+i\tilde\theta_-\tilde\psi^-}.
\end{gather}
Invariant vector f\/ields on Lie supergroups have to be treated with attention to signs to (see~\cite{C1}).
We def\/ine them by
\begin{gather*}
R_+g=\psi^+g, \qquad \tilde R_+g=\tilde\psi^+g, \qquad R_Xg=-Xg
\end{gather*}
for all $X$ in the complement of the $\psi^+$, $\tilde\psi^+$. This def\/inition guarantees that the invariant vector f\/ields obey the relations of~$\tgl$
\begin{gather*}
R_X R_Y- (-1)^{|X||Y|}R_YR_X=R_{[X, Y]}.
\end{gather*}
Then a computation reveals that
\begin{gather}
\psi^+g = -i\frac{d}{d\theta_+} g,\qquad \tilde\psi^+g = -i\frac{d}{d\tilde\theta_+} g, \qquad
Eg  = -i\frac{d}{dx} g,\qquad
\tilde Eg = -i\frac{d}{d\tilde x} g, \nonumber\\
Ng  = -i\frac{d}{dy} g+i\theta_+\psi^+g+i\tilde\theta_+\tilde\psi^+g,\qquad
\tilde Ng = -i\frac{d}{d\tilde y} g+i\theta_+\tilde\psi^+g, \label{eq:leftderiv}\\
\psi^-g = -ie^{iy} \frac{d}{d\theta_-} g+i\tilde y\tilde\psi^- g-i\theta_+Eg-i\tilde\theta_+\tilde Eg -\tilde y\theta_+\tilde Eg,\qquad
\tilde \psi^-g=-i e^{iy} \frac{d}{d\tilde\theta_-} g-i\theta_+\tilde Eg.\nonumber
\end{gather}
So that
\begin{gather*}
R_+ =  -i\frac{d}{d\theta_+},\qquad
\tilde R_+=  -i\frac{d}{d\tilde\theta_+},\qquad
R_E =  i\frac{d}{dx},\qquad
\tilde R_E=  i\frac{d}{d\tilde x}, \\ % \label{ra}\\
R_N  =  i\frac{d}{dy}-\theta_+ \frac{d}{d\theta_+}-\tilde\theta_+ \frac{d}{d\tilde\theta_+},\qquad
\tilde R_N =  i\frac{d}{d\tilde y}-\theta_+ \frac{d}{d\tilde\theta_+},\\
R_-  = ie^{iy} \left(\frac{d}{d\theta_-}+i\tilde y \frac{d}{d\tilde\theta_-}\right) -\theta_+\frac{d}{dx} -\tilde\theta_+\frac{d}{d\tilde x},\qquad
\tilde R_- = ie^{iy} \frac{d}{d\tilde\theta_-} -\theta_+\frac{d}{d\tilde x}.
\end{gather*}
In the same way for the left-action we require $L_-g=-g\psi^-$, $\tilde L_-g=-g\tilde\psi^-$ and $L_Xg=Xg$
for all $X$ in the complement of the $\psi^-$, $\tilde\psi^-$. Then one gets
\begin{gather*}
g\psi^-  = -i\frac{d}{d\theta_-} g,\qquad
 g\tilde\psi^- = -i\frac{d}{d\tilde\theta_-} g, \qquad
g E  = -i\frac{d}{dx} g,\qquad
 g\tilde E = -i\frac{d}{d\tilde x} g, \\
g N  = -i\frac{d}{dy} g+\theta_-\frac{d}{d\theta_-}g+\tilde\theta_-\frac{d}{d\tilde\theta_-}g, \qquad
g\tilde N = -i\frac{d}{d\tilde y} g+\theta_-\frac{d}{d\tilde\theta_-}g, \\
g\psi^+  = -ie^{iy} \left(\frac{d}{d\theta_+}+i\tilde y\frac{d}{d\tilde\theta_+}\right) g- \theta_-\frac{d}{dx}g-\tilde\theta_-\frac{d}{d\tilde x}g, \qquad
g\tilde\psi^+ = -ie^{iy} \frac{d}{d\tilde\theta_+}g-\theta_-\frac{d}{d\tilde x}g.
\end{gather*}
So that for the left-action we have
\begin{gather*}
L_-  = i\frac{d}{d\theta_-},\qquad
\tilde L_- = i\frac{d}{d\tilde\theta_-}, \qquad
L_E  = -i\frac{d}{dx},\qquad
\tilde L_E = -i\frac{d}{d\tilde x}, \\ % \label{la}\\
L_N  = -i\frac{d}{dy}+\theta_-\frac{d}{d\theta_-}+\tilde\theta_-\frac{d}{d\tilde\theta_-},\qquad
\tilde L_N = -i\frac{d}{d\tilde y}+\theta_-\frac{d}{d\tilde\theta_-}, \\
L_+  = -ie^{iy} \left(\frac{d}{d\theta_+}+i\tilde y\frac{d}{d\tilde\theta_+}\right)- \theta_-\frac{d}{dx}-\tilde\theta_-\frac{d}{d\tilde x},\qquad
\tilde L_+ = -ie^{iy} \frac{d}{d\tilde\theta_+}-\theta_-\frac{d}{d\tilde x}.
\end{gather*}

\subsection{The Haar measure}

The left-invariant Maurer--Cartan form is
\begin{gather*}
\omega(g) = g^{-1}dg,
\end{gather*}
and the right-invariant Maurer--Cartan form is $\omega(g^{-1})$. Either one can be taken to compute the Haar measure and we will use the right-invariant one
\begin{gather*}
\omega\big(g^{-1}\big) = -dg g^{-1}= -\left(\frac{d}{d\theta_+} g\right) g^{-1} d\theta_+ -\left(\frac{d}{d\tilde\theta_+} g\right) g^{-1} d\tilde\theta_+-\cdots .
\end{gather*}
This is a Lie superalgebra valued one form, that is it can be written as
\begin{gather*}
\omega\big(g^{-1}\big)  = \omega\big(\psi^+\big)\psi^+ + \omega\big(\tilde\psi^+\big)\tilde\psi^+ +\omega(E)E + \omega\big(\tilde E\big)\tilde E +
\omega(N)N + \omega\big(\tilde N\big)\tilde N \\
\hphantom{\omega\big(g^{-1}\big)  =}{}
+\omega\big(\psi^-\big)\psi^- + \omega\big(\tilde\psi^-\big)\tilde\psi^-.
\end{gather*}
The components are the dual one-forms in our basis.
Using \eqref{eq:leftderiv} they can be easily extracted
\begin{alignat*}{3}
& \omega\big(\psi^+\big) = -id\theta_+-\theta_+dy, \qquad &&  \omega\big(\tilde\psi^+\big)=-id\tilde\theta_+-\tilde\theta_+dy-\theta_+d\tilde y, &\\
& \omega(E) = -idx+e^{-iy}\theta_+d\theta_-,\qquad  && \omega\big(\tilde E\big)= -id\tilde x +e^{-iy}\big(\tilde\theta_+-i\tilde y \theta_+\big)d\theta_-+e^{-iy}\theta_+d\tilde\theta_-, & \\
& \omega(N) = -idy,\qquad &&  \omega(\tilde N)=-id\tilde y, & \\
& \omega\big(\psi^-\big) = -ie^{-iy}d\theta_-,\qquad &&  \omega\big(\tilde\psi^-\big)=-ie^{-iy}d\tilde\theta_--e^{-iy}\tilde yd\theta_-. &
\end{alignat*}
The right-invariant measure is the wedge product of these dual one forms
\begin{gather*}
\mu\big(g^{-1}\big)  = \omega\big(\psi^+\big)\wedge\omega\big(\tilde\psi^+\big)\wedge \omega(E)\wedge\omega\big(\tilde E\big)\wedge \omega(N)\wedge\omega\big(\tilde N\big)\wedge \omega\big(\psi^-\big)\wedge\omega\big(\tilde\psi^-\big)\\
\hphantom{\mu\big(g^{-1}\big)}{}
 = e^{-2iy} d\theta_+\wedge d\tilde\theta_+\wedge dx\wedge d\tilde x \wedge dy\wedge d\tilde y \wedge d\theta_-\wedge d\tilde\theta_-.
\end{gather*}
Here, we used graded anti-symmetry of the wedge product as well as integration with respect to an odd variable is the same as taking the derivative, but the double derivative of
an odd variable vanishes, in other words we have used $d\theta\wedge d\theta =0$.

\subsection{The Laplace  operators}

In~\cite{BR} two Casimir operators of $\tgl$ were given. We change one of them by the central ele\-ment~$\tilde E$. They are then
\begin{gather*}
Q_1 = N\tilde E+\tilde N E+\psi^-\tilde\psi^++\tilde \psi^-\psi^+-\tilde E, \qquad
Q_2 =\tilde N\tilde E+\tilde\psi^-\tilde\psi^+.
\end{gather*}
We compute that
\begin{gather*}
\Delta_1:=Q_1^{\rm L}  = Q_1^{\rm R} = -\frac{d}{d y}\frac{d}{d\tilde x}-i\frac{d}{d\tilde x} -\frac{d}{d\tilde y}\frac{d}{d x} +
e^{iy}\left(\frac{d}{d\theta_-}\frac{d}{d\tilde\theta_+}+\frac{d}{d\tilde\theta_-}\frac{d}{d\theta_+}+i\tilde y\frac{d}{d\tilde\theta_-}\frac{d}{d\tilde\theta_+} \right),\\
\Delta_2:=Q_2^{\rm L}  = Q_2^{\rm R} = -\frac{d}{d \tilde y}\frac{d}{d\tilde x} +e^{iy}\frac{d}{d\tilde\theta_-}\frac{d}{d\tilde\theta_+}.
\end{gather*}
We call these operators $\Delta$ as they are the Laplace operators on this supergroup.
This f\/inishes our preparations, and we can turn to decomposing functions with respect to the left-right action of invariant vector f\/ields.

\subsection{Decomposing functions}

Before we start decomposing functions, let us observe how left and right action are related,

\begin{Proposition}\label{prop:map}
The change of coordinates
\begin{gather*}
\beta\colon \  \theta_\pm \mapsto -\theta_\mp, \qquad  \tilde\theta_\pm \mapsto -\tilde\theta_\mp,\qquad x \mapsto -x, \qquad \tilde x \mapsto -\tilde x, \qquad
y\mapsto y, \qquad \tilde y \mapsto \tilde y
\end{gather*}
relates the action of left- and right-invariant vector fields as
\begin{gather*}
R_\pm\mapsto L_\mp, \qquad \tilde R_\pm \mapsto \tilde L_\mp, \qquad R_E\mapsto L_E, \qquad \tilde R_E \mapsto \tilde L_E, \\ R_N \mapsto -L_N, \qquad \tilde R_N \mapsto -\tilde L_N.
\end{gather*}
\end{Proposition}
It will turn out that there are three-types of modules appearing in the decomposition. We will call these modules typical, semi-typical and atypical as in~\cite{BR}.

\subsubsection{Typical modules}

Let
\begin{gather*}
f_{e,n,\tilde{e},\tilde{n}}(x,\tilde{x},y,\tilde{y})=\exp(ixe+i\tilde{x}\tilde{e}-iyn-i\tilde{n}\tilde{y}),
\end{gather*}
then the crucial lemma is:
\begin{Lemma}\label{lem:typ}
Define the matrix
\begin{gather*}
M=\begin{pmatrix}  1 & \theta_- & \tilde\theta_- & \theta_-\tilde\theta_- \\
\theta_+ &   \theta_+\theta_- & \theta_+\tilde \theta_- & \theta_+\theta_-\tilde\theta_- \\
\tilde\theta_+ & \tilde\theta_+\theta_- & \tilde\theta_+\tilde\theta_- & \tilde \theta_+\theta_-\tilde\theta_- \\
\theta_+\tilde\theta_+ &\theta_+\tilde \theta_+\theta_- & \theta_+\tilde\theta_+\tilde\theta_- & \theta_+\tilde\theta_+ \theta_-\tilde\theta_-
\end{pmatrix}
\end{gather*}
and the second-order differential operator
\begin{gather*}
\mathcal D_{e, \tilde e} = \frac{d}{d\tilde\theta_-} \frac{d}{d\theta_+} + \frac{d}{d\theta_-}\frac{d}{d\tilde\theta_+} +\left( i\tilde y -\frac{e}{\tilde e}\right) \frac{d}{d\tilde\theta_-} \frac{d}{d\tilde\theta_+}.
\end{gather*}
Let $\tilde e\neq 0$, then each row of the matrix
\begin{gather*}
V_{e, n, \tilde e, \tilde n}  = f_{e, n, \tilde e, \tilde n}\text{exp} \left( \frac{e^{iy}}{\tilde e} \mathcal D_{e, \tilde e} \right) M
\end{gather*}
carries the irreducible highest-weight representation of the left-regular action of highest-weight $(e, \tilde e, -n+2, -\tilde n)$. Each column transforms in the
 irreducible highest-weight representation of the right-regular action of highest-weight $(-e, -\tilde e, n, \tilde n)$.
\end{Lemma}

\begin{proof}
We def\/ine the function $f_4=\theta_-\tilde{\theta}_-f_{e,n,\tilde{e},\tilde{n}}$, then$f_4$ is a highest-weight vector for both left- and right-action. The weight under the right-action is
$\left( -e, -\tilde e, n, \tilde n\right)$ and the weight under the left-action is $\left(e, \tilde e, -n+2, -\tilde n\right)$.
Then under the left-action, the functions
\begin{gather*}
f_1:=f_{e,n,\tilde{e},\tilde{n}}(x,\tilde{x},y,\tilde{y}), \qquad
f_2:=\theta_- f_{e,n,\tilde{e},\tilde{n}}, \qquad f_3:=\tilde{\theta}_-f_{e,n,\tilde{e},\tilde{n}}, \qquad
f_4:=\theta_-\tilde{\theta}_-f_{e,n,\tilde{e},\tilde{n}},
\end{gather*}
carry the four-dimensional irreducible representation of that highest-weight.
Since left- and right-action commute, each of these states must be a highest-weight vector of same weight as before for the right-action. For the highest-weight $f_1$ the remaining three states of the typical module~I are
\begin{gather*}
\begin{split}
&\tilde{\psi}_{\rm I} =\tilde{R}_-f_1=i\tilde{e}\theta_+ f_{e,n,\tilde{e},\tilde{n}},\qquad % \label{rI}\\
\psi_{\rm I} =R_-f_1=ie\theta_+ f_{e,n,\tilde{e},\tilde{n}}+i\tilde{e}\tilde{\theta}_+ f_{e,n,\tilde{e},\tilde{n}},\\
& b_{\rm I}=R_-\tilde{R}_-f_1=-\tilde{e}^2\theta_+\tilde{\theta}_+f_{e,n,\tilde{e},\tilde{n}}.
\end{split}
\end{gather*}
In the same way three states completing the module~II with the highest-weight~$f_2$ are
\begin{gather*}
\tilde{\psi}_{\rm II} =\tilde{R}_-f_2=-i\tilde{e}\theta_-\theta_+f_{e,n,\tilde{e},\tilde{n}},\\ %\label{rII}\\
\psi_{\rm II} =R_-f_2=if_{e,n-1,\tilde{e},\tilde{n}}-ie\theta_-\theta_+f_{e,n,\tilde{e},\tilde{n}}
-i\tilde{e}\theta_-\tilde{\theta}_+f_{e,n,\tilde{e},\tilde{n}},\\
b_{\rm II} =R_-\tilde{R}_-f_2=\tilde{e}\theta_+f_{e,n-1,\tilde{e},\tilde{n}}+
\tilde{e}^2\theta_-\theta_+\tilde{\theta}_+f_{e,n,\tilde{e},\tilde{n}}.
\end{gather*}
For the module~III with the highest-weight $f_3$ we get
\begin{gather*}
\tilde{\psi}_{\rm III} =\tilde{R}_-f_3=if_{e,n-1,\tilde{e},\tilde{n}}
+i\tilde{e}f_{e,n,\tilde{e},\tilde{n}}\theta_+\tilde{\theta}_-, \\ %\label{rIII}\\
\psi_{\rm III} =R_-f_3=-\tilde{y}f_{e,n-1,\tilde{e},\tilde{n}}-ief_{e,n,\tilde{e},\tilde{n}}\tilde{\theta}_-\theta_+
-i\tilde{e}f_{e,n,\tilde{e},\tilde{n}}\tilde{\theta}_-\tilde{\theta}_+ ,\\
b_{\rm III} =R_-\tilde{R}_-f_3=-ef_{e,n-1,\tilde{e},\tilde{n}}\theta_+-\tilde{e}f_{e,n-1,\tilde{e},\tilde{n}}
\tilde{\theta}_++i\tilde{e}\tilde{y}f_{e,n-1,\tilde{e},\tilde{n}}\theta_++\tilde{e}^2f_{e,n,\tilde{e},\tilde{n}}
\tilde{\theta}_-\theta_+\tilde{\theta}_+,
\end{gather*}
and for the last module IV with highest-weight $f_4$
\begin{gather*}
\tilde{\psi}_{\rm IV} =\tilde{R}_-f_4=-if_{e,n-1,\tilde{e},\tilde{n}}\theta_-
+i\tilde{e}f_{e,n,\tilde{e},\tilde{n}}\theta_-\tilde{\theta}_-\theta_+, \\ % \label{rIV}\\
\psi_{\rm IV} =R_-f_4=\tilde{y}f_{e,n-1,\tilde{e},\tilde{n}}\theta_-+if_{e,n-1,\tilde{e},\tilde{n}}\tilde{\theta}_-
+ief_{e,n,\tilde{e},\tilde{n}}\theta_-\tilde{\theta}_-\theta_++i\tilde{e}f_{e,n,\tilde{e},\tilde{n}}
\theta_-\tilde{\theta}_-\tilde{\theta}_+ ,\\
b_{\rm IV} =R_-\tilde{R}_-f_4=f_{e,n-2,\tilde{e},\tilde{n}}-\tilde{e}f_{e,n-1,\tilde{e},\tilde{n}}\tilde{\theta}_-\theta_+
+(i\tilde{y}\tilde{e}-e)f_{e,n-1,\tilde{e},\tilde{n}}\theta_-\theta_+\\
\hphantom{b_{\rm IV} =}{} +\tilde{e}f_{e,n-1,\tilde{e},\tilde{n}}
\tilde{\theta}_+\theta_-+\tilde{e}^2f_{e,n,\tilde{e},\tilde{n}}\theta_-\tilde{\theta}_-\theta_+\tilde{\theta}_+.
\end{gather*}

The action of $R_N$ is the same on all four four modules:
\begin{gather*}
R_N f_i= n f_i, \qquad R_N \psi = (n-1)\psi, \qquad R_N \tilde{\psi} = (n-1)\tilde{\psi}, \qquad R_N b = (n-2)b.
\end{gather*}
The action of $\tilde{R}_N$ is diagonal on the states $X=f_i,\tilde{\psi},b$: $\tilde{R}_N X=\tilde{n} X$, but non diagonal on the states $\psi$:
\begin{gather*}
\tilde{R}_N \psi=\tilde{n}\psi-\tilde{\psi}.
\end{gather*}

In summary, for $\tilde e\neq 0$,
each row of the matrix
\begin{gather*}
  f_{e, n, \tilde e, \tilde n} \begin{pmatrix}  1 & \theta_- & \tilde\theta_- & \theta_-\tilde\theta_- \\
\tilde e \theta_+ &  \tilde e \theta_-\theta_+ & \tilde e \theta_+\tilde \theta_- & -\tilde e \theta_-\tilde\theta_-\theta_+ \\
e\theta_+ +\tilde e \tilde\theta_+ & -e\theta_-\theta_+ -\tilde e \theta_-\tilde\theta_+ & e\tilde\theta_-\theta_+ +\tilde e \tilde\theta_-\tilde\theta_+ & e\theta_-\tilde\theta_-\theta_+ +\tilde e \theta_-\tilde\theta_-\tilde\theta_+ \\
\tilde e^2 \theta_+\tilde\theta_+ & \tilde e^2 \theta_-\theta_+\tilde \theta_+ & \tilde e^2 \tilde\theta_-\theta_+\tilde\theta_+ & \tilde e^2 \theta_-\tilde\theta_- \theta_+\tilde\theta_+
\end{pmatrix}   \\
 \qquad {} + f_{e, n-1, \tilde e, \tilde n} \begin{pmatrix}  0 & 0 & 0 & 0 \\
0 &  0 & 1  & \theta_-  \\
0 & 1 & -i\tilde y   & -i\tilde y  \theta_- +\tilde\theta_- \\
0 & \tilde e \theta_+  & -e\theta_+ -\tilde e  \tilde\theta_+ +i\tilde e\tilde y \theta_+  & -\tilde e\tilde\theta_-\theta_+ +(i\tilde y \tilde e-e) \theta_-\theta_+ +\tilde e \tilde\theta_+\theta_-
\end{pmatrix} \\
 \qquad {}+ f_{e, n-2, \tilde e, \tilde n} \begin{pmatrix}  0 & 0 & 0 & 0 \\
0 &  0 & 0  & 0  \\
0 & 0 & 0  & 0 \\
0 & 0 & 0 & 1
\end{pmatrix}
\end{gather*}
carries the irreducible highest-weight representation of the left regular action of highest-weight $(e, \tilde e, -n+2, -\tilde n)$. Each column transforms in the
 irreducible highest-weight representation of the right regular action of highest-weight $(-e, -\tilde e, n, \tilde n)$.
Changing basis $\psi\rightarrow \psi -\frac{e}{\tilde e}\tilde \psi$, this can be written more symmetrically,
\begin{gather*}
 f_{e, n, \tilde e, \tilde n} \begin{pmatrix}  1 & \theta_- & \tilde\theta_- & \theta_-\tilde\theta_- \\
\theta_+ &   \theta_+\theta_- & \theta_+\tilde \theta_- & \theta_+\theta_-\tilde\theta_- \\
\tilde\theta_+ & \tilde\theta_+\theta_- & \tilde\theta_+\tilde\theta_- & \tilde \theta_+\theta_-\tilde\theta_- \\
\theta_+\tilde\theta_+ &\theta_+\tilde \theta_+\theta_- & \theta_+\tilde\theta_+\tilde\theta_- & \theta_+\tilde\theta_+ \theta_-\tilde\theta_-
\end{pmatrix}   \\
\qquad{}+ \frac{f_{e, n-1, \tilde e, \tilde n}}{\tilde e} \begin{pmatrix}  0 & 0 & 0 & 0 \\
0 &  0 & 1  & -\theta_-  \\
0 & 1 & \left(i\tilde y-\frac{e}{\tilde e}\right)   & -\theta_-\left(i\tilde y-\frac{e}{\tilde e}\right)+\tilde \theta_-\\
0 &  \theta_+  & \theta_+\left(i\tilde y-\frac{e}{\tilde e}\right) - \tilde\theta_+  & \left(i\tilde y-\frac{e}{\tilde e}\right)\theta_-\theta_+ - \tilde\theta_-\theta_+ +\tilde\theta_+\theta_-
\end{pmatrix}\\
\qquad {} +
\frac{f_{e, n-2, \tilde e, \tilde n}}{\tilde e^{2}} \begin{pmatrix}  0 & 0 & 0 & 0 \\
0 &  0 & 0  & 0  \\
0 & 0 & 0  & 0 \\
0 & 0 & 0 & 1
\end{pmatrix},
\end{gather*}
so that the lemma follows.
\end{proof}

\begin{Remark}
The operator $e^{iy}\mathcal D_{e, \tilde e}$ is the semi-classical analouge of a screening charge. It is related to the Casimir operators acting on $V_{e, n, \tilde e, \tilde n}$ via
\begin{gather*}
\frac{1}{\tilde e}\left( \Delta_1-\frac{e}{\tilde e} \Delta_2 -e^{iy}\mathcal D_{e, \tilde e}\right) = -i\frac{d}{dy} +1.
\end{gather*}
In other words the dif\/ference of the Casimir $\Delta_1-\frac{e}{\tilde e} \Delta_2$ and $e^{iy}\mathcal D_{e, \tilde e}$ is a semi-simple operator.
\end{Remark}
Finally, we def\/ine the matrices
\begin{gather*}
V_{e, n, \tilde e, \tilde n}^{(m)} := \tilde y^m V_{e, n, \tilde e, \tilde n},
\end{gather*}
and denote the vector space spanned by its components by $T_{e, n, \tilde e, \tilde n}^{(m)}$, so that
\begin{gather*}
T_{e, n, \tilde e, \tilde n} := \bigoplus_{m=0}^\infty T_{e, n, \tilde e, \tilde n}^{(m)}
\end{gather*}
is an inf\/inite-dimensional module for the left-right action of $\tgl$.
The irredicible submodule $T_{e, n, \tilde e, \tilde n}^{(0)}$ has been described in Lemma~\ref{lem:typ}. The modules $\bigoplus\limits_{m=0}^N T_{e, n, \tilde e, \tilde n}^{(m)}$ are all submodules with quotient by $\bigoplus\limits_{m=0}^{N-1} T_{e, n, \tilde e, \tilde n}^{(m)}$ isomorphic to $T_{e, n, \tilde e, \tilde n}^{(0)}$.

\subsubsection{Semitypical modules}

The case $\tilde e=0$ and $e\neq 0$ is called semi-typical. Our strategy is to f\/irst
decompose into irreducible $\gl$ modules.
First, f\/ix $e\neq 0$, $\tilde n$. In order to avoid too many indices we take the short-hand notation $f_n:= f_{e, n, 0, \tilde n}$.
We f\/ind the following list for the left-action~$L_\pm$ on weight modules:
\begin{gather*}
A_{1, n}  = \text{span} (f_n, \theta_- f_n   ),\\
A_{2, n}  = \text{span} (\theta_+f_n, ie\theta_-\theta_+ f_n-if_{n-1}   ),\\
A_{3, n}  = \text{span}\big(\tilde\theta_+f_n, ie\theta_-\tilde\theta_+ f_n+\tilde y f_{n-1}  \big),\\
A_{4, n}  = \text{span}\big(\tilde\theta_-f_n, \theta_-\tilde\theta_- f_n  \big),\\
A_{5, n}  = \text{span}\big(\theta_+\tilde\theta_+f_n, ie\theta_-\theta_+\tilde\theta_+ f_n-i\tilde\theta_+f_{n-1}-\tilde y \theta_+ f_{n-1}  \big),\\
A_{6, n}  = \text{span}\big(\tilde\theta_-\tilde\theta_+f_n, ie\tilde \theta_-\theta_-\tilde\theta_+ f_n+\tilde y\tilde\theta_-f_{n-1}  \big),\\
A_{7, n}  = \text{span}\big(\tilde\theta_-\theta_+f_n, ie\tilde\theta_-\theta_-\theta_+ f_n-i\tilde\theta_-f_{n-1}  \big),\\
A_{8, n}  = \text{span}\big(\tilde\theta_-\theta_+\tilde\theta_+f_n, ie\tilde\theta_-\theta_-\theta_+\tilde\theta_+ f_n-i\tilde\theta_-\tilde\theta_+f_{n-1}-\tilde y \tilde\theta_-\theta_+f_{n-1}  \big).
\end{gather*}
We then also def\/ine for non-negative integer $m$
\begin{gather*}
A_{i,n}^{(m)}:= \tilde y^m A_{i, n}.
\end{gather*}
The $\tilde L_\pm$ action gives $\gl$-module one-to-one maps, that only change the $L_N$-action,
\begin{alignat*}{3}
&\tilde L_+\colon \  A_{3, n}^{(m)} \rightarrow A_{1, n-1}^{(m)}, \qquad && \tilde L_-\colon \  A_{4, n}^{(m)} \rightarrow A_{1, n}^{(m)},&\\
&\tilde L_+\colon \  A_{5, n}^{(m)} \rightarrow A_{2, n-1}^{(m)}, \qquad &&\tilde L_-\colon \  A_{6, n}^{(m)} \rightarrow A_{3, n}^{(m)},&\\
&\tilde L_+\colon \  A_{6, n}^{(m)} \rightarrow A_{4, n-1}^{(m)}, \qquad &&\tilde L_-\colon \  A_{7, n}^{(m)} \rightarrow A_{2, n}^{(m)},&\\
&\tilde L_+\colon \  A_{8, n}^{(m)} \rightarrow A_{7, n-1}^{(m)}, \qquad &&\tilde L_-\colon \  A_{8, n}^{(m)} \rightarrow A_{5, n}^{(m)}.&
\end{alignat*}
The $\tilde L_N$-action on $\gl$-modules is described as follows. $\tilde L_N$ splits into a semi-simple and a~nilpotent part. The semi-simple part acts by multiplication by~$-\tilde n$, while the nilpotent part acts as follows
\begin{gather*}
{\tilde L_N}^{\text{\rm nil}}\colon \   A_{i,n}^{(m)}  \rightarrow A_{i, n}^{(m-1)}\oplus \begin{cases}
0 & \text{if} \ i = 1,2, \\
A_{1, n-1}^{(m)} & \text{if} \ i =3,\\
A_{1, n}^{(m)} & \text{if} \ i =4,\\
A_{2, n-1}^{(m)} & \text{if} \ i =5,\\
A_{3,n}^{(m)} \oplus A_{4,n-1}^{(m)} \oplus A_{1,n-1}^{(m+1)} &  \text{if} \ i =6,\\
A_{1,n-1}^{(m)}\oplus A_{2,n}^{(m)} & \text{if} \ i =7,\\
A_{5,n}^{(m)}\oplus A_{7,n-1}^{(m)}\oplus A_{3,n-1}^{(m)}\oplus A_{2, n-1}^{(m+1)}\oplus A_{1, n-2}^{(m+1)}&  \text{if} \ i =8.
\end{cases}
\end{gather*}
By this we mean that the $\gl$ action on the image of the nilpotent part is the right-hand side. We also understand $A^{(-1)}_{i,n}=0$.
An ef\/f\/icient way to picture these computations is the following

\begin{Proposition}
Let $\tilde e=0$ and $e\neq 0$, then under the left-regular action as $\tsl$ modules, we have the following two types of $8$-dimensional $\tsl$-modules,
$S^{(m)}_{n, a}$
\begin{gather*}
\begin{tikzpicture}[auto,thick]
\node (top) at (0,1.5) [] {$A^{(m)}_{4,n-1}$};
\node (left) at (-1.5,0) [] {$A^{(m)}_{6,n}$};
\node (right) at (1.5,0) [] {$A^{(m)}_{1,n-1}$};
\node (bot) at (0,-1.5) [] {$A^{(m)}_{3,n}$};
\draw [->] (left) to (top);
\draw [->] (left) to (bot);
\draw [->] (top) to (right);
\draw [->] (bot) to (right);
%\draw [->,dotted] (0.2,1.2) arc [radius=1.5, start angle=190,end angle=240];
%\draw [->,dotted] (-1.3,-0.1) arc [radius=1.5, start angle=80,end angle=15];
%\node at (0.35,0.35) [] {$\tilde{L}_N$};
%\node at (-0.2,-0.2) [] {$\tilde{L}_N$};
\node at (-0.95,-0.95) [] {$\tilde{L}_-$};
\node at (1,1) [] {$\tilde{L}_-$};
\node at (-1,1) [] {$\tilde{L}_+$};
\node at (1,-1) [] {$\tilde{L}_+$};
\end{tikzpicture}
\end{gather*}
and $S^{(m)}_{n, b}$
\begin{gather*}
\begin{tikzpicture}[auto,thick]
\node (top) at (0,1.5) [] {$A^{(m)}_{7,n-1}$};
\node (left) at (-1.5,0) [] {$A^{(m)}_{8,n}$};
\node (right) at (1.5,0) [] {$A^{(m)}_{2,n-1}$};
\node (bot) at (0,-1.5) [] {$A^{(m)}_{5,n}$};
\draw [->] (left) to (top);
\draw [->] (left) to (bot);
\draw [->] (top) to (right);
\draw [->] (bot) to (right);
%\draw [->,dotted] (0.2,1.2) arc [radius=1.5, start angle=190,end angle=240];
%\draw [->,dotted] (-1.3,-0.1) arc [radius=1.5, start angle=80,end angle=15];
%\node at (0.35,0.35) [] {$\tilde{L}_N$};
%\node at (-0.2,-0.2) [] {$\tilde{L}_N$};
\node at (-0.95,-0.95) [] {$\tilde{L}_-$};
\node at (1,1) [] {$\tilde{L}_-$};
\node at (-1,1) [] {$\tilde{L}_+$};
\node at (1,-1) [] {$\tilde{L}_+$};
\end{tikzpicture}
\end{gather*}
While $L_N$-acts semi-simple on these modules, the nilpotent part of~$\tilde L_N$ acts as
\begin{gather*}
 {\tilde L_N}^{\text{\rm nil}}\colon \  S^{(m)}_{n, a} \rightarrow S^{(m-1)}_{n, a} \oplus S^{(m)}_{n, a}\oplus S^{(m+1)}_{n, a}, \\
 {\tilde L_N}^{\text{\rm nil}}\colon \  S^{(m)}_{n, b} \rightarrow S^{(m-1)}_{n, b} \oplus S^{(m)}_{n, b}\oplus S^{(m+1)}_{n, b}\oplus S^{(m)}_{n, a}\oplus S^{(m+1)}_{n, a}.
\end{gather*}
\end{Proposition}

However, with a suitable change of basis the situation can be improved. Def\/ine
\begin{gather*}
\check{A}^{(m)}_{6,n}  := \text{span}\left(\tilde y^m\left(\frac{1}{2} \tilde y^2 f_{n-1}-e f_n \tilde\theta_-\tilde\theta_+ , \; e\tilde y f_{n-1} \tilde\theta_- +\frac{i}{2}\tilde y^2 f_{n-1}\theta_- -i e f_n \theta_-\tilde\theta_-\tilde\theta_+\right)\right),\\
\check{A}^{(m)}_{8,n}  := \text{span}\left(\tilde y^m\left(ie f_n\tilde\theta_-\theta_+\tilde\theta_+ +\tilde y f_{n-1}\tilde\theta_+ -\frac{i}{2} \tilde y^2 f_{n-1}\theta_+,\;\frac{1}{2}\tilde y^2 f_{n-2} \right.\right. \\
 \left.\left. \hphantom{\check{A}^{(m)}_{8,n}  := }{}
 +\frac{e}{2}\tilde y^2 f_{n-1}\theta_-\theta_+ +i e \tilde y f_{n-1}(\tilde\theta_-\theta_+ +\theta_-\tilde\theta_+)-e f_{n-1}\tilde\theta_-\tilde\theta_+ -e^2 f_{n}\theta_-\tilde\theta_-\theta_+\tilde\theta_+\right)\right),\\
 \check{A}^{(m)}_{7,n}  :=\text{span}\left(\tilde y^m\left(e f_{n}\tilde\theta_-\theta_+ -i \tilde y f_{n-1}, \; \tilde y f_{n-1}\theta_- +i f_{n-1}\tilde\theta_- +ie\theta_-\tilde\theta_-\theta_+\right)\right), \\
\check{A}^{(m)}_{i,n}  := A^{(m)}_{i,n} \qquad  \text{for}  \ i=1, 2, 3,  4, 5.
\end{gather*}
This redef\/inition preserves the $\tsl$ action.
However the nilpotent part of $\tilde L_N$ acts as
\begin{gather}\label{eq:Ltildenil}
{\tilde L_N}^{\text{nil}}\colon \  \check A_{i,n}^{(m)}  \rightarrow\check A_{i, n}^{(m-1)}\oplus \begin{cases}
0 & \text{if} \ i = 1,2, \\
\check A_{1, n-1}^{(m)} & \text{if} \ i =3,\\
\check A_{1, n}^{(m)} & \text{if} \ i =4,\\
\check A_{2, n-1}^{(m)} & \text{if} \ i =5,\\
\check A_{3,n}^{(m)}  \oplus\check A_{1,n-1}^{(m)} &  \text{if} \ i =6,\\
 \check A_{2,n}^{(m)} &  \text{if} \ i =7,\\
\check A_{5,n}^{(m)}\oplus \check A_{2, n-1}^{(m)}&  \text{if} \ i =8.
\end{cases}
\end{gather}
We thus can considerably improve the previous proposition.
\begin{Theorem}
Let $\tilde e=0$ and $e\neq 0$, then under the left-regular action of $\tgl$, we have the following modules,
$\check S^{(m)}_{n, a}$
\begin{gather*}
\begin{tikzpicture}[auto,thick]
\node (top) at (0,1.5) [] {$\check A^{(m)}_{4,n-1}$};
\node (left) at (-1.5,0) [] {$\check A^{(m)}_{6,n}$};
\node (right) at (1.5,0) [] {$\check A^{(m)}_{1,n-1}$};
\node (bot) at (0,-1.5) [] {$\check A^{(m)}_{3,n}$};
\draw [->] (left) to (top);
\draw [->] (left) to (bot);
\draw [->] (top) to (right);
\draw [->] (bot) to (right);
%\draw [->,dotted] (0.2,1.2) arc [radius=1.5, start angle=190,end angle=240];
%\draw [->,dotted] (-1.3,-0.1) arc [radius=1.5, start angle=80,end angle=15];
%\node at (0.35,0.35) [] {$\tilde{L}_N$};
%\node at (-0.2,-0.2) [] {$\tilde{L}_N$};
\node at (-0.95,-0.95) [] {$\tilde{L}_-$};
\node at (1,1) [] {$\tilde{L}_-$};
\node at (-1,1) [] {$\tilde{L}_+$};
\node at (1,-1) [] {$\tilde{L}_+$};
\end{tikzpicture}
\end{gather*}
and $\check S^{(m)}_{n, b}$
\begin{gather*}
\begin{tikzpicture}[auto,thick]
\node (top) at (0,1.5) [] {$\check A^{(m)}_{7,n-1}$};
\node (left) at (-1.5,0) [] {$\check A^{(m)}_{8,n}$};
\node (right) at (1.5,0) [] {$\check A^{(m)}_{2,n-1}$};
\node (bot) at (0,-1.5) [] {$\check A^{(m)}_{5,n}$};
\draw [->] (left) to (top);
\draw [->] (left) to (bot);
\draw [->] (top) to (right);
\draw [->] (bot) to (right);
%\draw [->,dotted] (0.2,1.2) arc [radius=1.5, start angle=190,end angle=240];
%\draw [->,dotted] (-1.3,-0.1) arc [radius=1.5, start angle=80,end angle=15];
%\node at (0.35,0.35) [] {$\tilde{L}_N$};
%\node at (-0.2,-0.2) [] {$\tilde{L}_N$};
\node at (-0.95,-0.95) [] {$\tilde{L}_-$};
\node at (1,1) [] {$\tilde{L}_-$};
\node at (-1,1) [] {$\tilde{L}_+$};
\node at (1,-1) [] {$\tilde{L}_+$};
\end{tikzpicture}
\end{gather*}
While $L_N$-acts semi-simple on these modules, the nilpotent part of $\tilde L_N$ acts as
described in equation \eqref{eq:Ltildenil}. We can split it into two pieces ${\tilde L_N}^{\text{\rm nil}}= \tilde N_a+\tilde N_b$, such that
\begin{gather*}
\tilde N_a\colon \   \check A_{i,n}^{(m)} \rightarrow\check A_{i, n}^{(m-1)}
\end{gather*}
and
\begin{gather*}
\tilde N_b \colon \   \check A_{i,n}^{(m)} \rightarrow
\begin{cases}
0 &  \text{if} \ i = 1,2, \\
\check A_{1, n-1}^{(m)} & \text{if} \ i =3,\\
\check A_{1, n}^{(m)} & \text{if} \ i =4,\\
\check A_{2, n-1}^{(m)} & \text{if} \ i =5,\\
\check A_{3,n}^{(m)}  \oplus\check A_{1,n-1}^{(m)} &  \text{if} \ i =6,\\
 \check A_{2,n}^{(m)} & \text{if} \ i =7,\\
\check A_{5,n}^{(m)}\oplus \check A_{2, n-1}^{(m)}& \text{if} \ i =8.
\end{cases}
\end{gather*}
The long-exact sequence describes the action of $\tilde N_a$,
\begin{gather*}
\cdots \longrightarrow \check S^{(m)}_{n, c} \longrightarrow \check S^{(m-1)}_{n, c} \longrightarrow \dots\longrightarrow \check S^{(1)}_{n, c} \longrightarrow\check S^{(0)}_{n, c} \longrightarrow 0
\end{gather*}
for $c\in\{ a, b\}$.
\end{Theorem}
In order to understand the left-right action, we f\/irst have to understand the left-right $\gl$ action.
\begin{Definition}
The highest-weight representation of $\gl$, with highest-weight $(e, n)$ and lowest-weight $(e,n-1)$ will be denoted by $\rho_{e, n-\frac{1}{2}}$.
\end{Definition}

We compute: Under the left-right $\gl$ action the modules decompose as
\begin{gather*}
X^{(m)}_{2, n}:=\check A^{(m)}_{6, n} \oplus \check A^{(m)}_{8, n}  = \rho^L_{e,-n+\frac{3}{2}}\otimes  \rho^R_{-e,n-\frac{3}{2}}, \\
X^{(m)}_{-, n}:=\check A^{(m)}_{3, n} \oplus \check A^{(m)}_{5, n}  = \rho^L_{e,-n+\frac{1}{2}}\otimes  \rho^R_{-e,n-\frac{3}{2}}, \\
X^{(m)}_{+, n-1}:=\check A^{(m)}_{4, n-1} \oplus \check A^{(m)}_{7, n-1}  = \rho^L_{e,-n+\frac{5}{2}}\otimes  \rho^R_{-e,n-\frac{3}{2}}, \\
X^{(m)}_{0, n-1}:=\check A^{(m)}_{1, n-1} \oplus \check A^{(m)}_{2, n-1}  = \rho^L_{e,-n+\frac{3}{2}}\otimes  \rho^R_{-e,n-\frac{3}{2}},
\end{gather*}
where the last three decompositions follow from the f\/irst one and the f\/irst one is a straight-forward computation.
Further, the $\tilde R_\pm$ action gives $\gl$-module one-to-one maps, that only change the $R_N$-action,
\begin{alignat*}{3}
 &\tilde R_- \colon \  A_{4, n}^{(m)} \rightarrow A_{1, n-1}^{(m)}, \qquad &&\tilde R_+ \colon \  A_{3, n}^{(m)} \rightarrow A_{1, n}^{(m)},&\\
 &\tilde R_- \colon \ A_{6, n}^{(m)} \rightarrow A_{3, n-1}^{(m)}, \qquad &&\tilde R_+ \colon \  A_{5, n}^{(m)} \rightarrow A_{2, n}^{(m)},&\\
 &\tilde R_- \colon \ A_{7, n}^{(m)} \rightarrow A_{2, n-1}^{(m)}, \qquad &&\tilde R_+ \colon \  A_{6, n}^{(m)} \rightarrow A_{4, n}^{(m)},&\\
 &\tilde R_- \colon \ A_{8, n}^{(m)} \rightarrow A_{5, n-1}^{(m)}, \qquad &&\tilde R_+ \colon \  A_{8, n}^{(m)} \rightarrow A_{7, n}^{(m)}.&
\end{alignat*}
We thus get the picture for left-right modules given in Fig.~\ref{fig:semtyp}.
\begin{figure}[t]
\centering
\begin{tikzpicture}[thick,>=latex,scale=1.4]
\node (tr) at (12,4) [] {$\scriptstyle X^{(m)}_{2, n+1}$};
\node (tm) at (8,4) [] {$\scriptstyle X^{(m)}_{2, n}$};
\node (tl) at (4,4) [] {$\scriptstyle X^{(m)}_{2, n-1}$};
\node (br) at (12,2) [] {$\scriptstyle X^{(m)}_{0, n}$};
\node (bm) at (8,2) [] {$\scriptstyle X^{(m)}_{0, n-1}$};
\node (bl) at (4,2) [] {$\scriptstyle X^{(m)}_{0, n-2}$};
\node at (13,4) [] {$\cdots$};
\node (mrr) at (13,3) [] {$\cdots$};
\node at (13,2) [] {$\cdots$};
\node (mrl) at (11,3) [] {$\scriptstyle X^{(m)}_{+, n}$};
\node (mmr) at (9,3) [] {$\scriptstyle X^{(m)}_{-, n}$};
\node (mml) at (7,3) [] {$\scriptstyle X^{(m)}_{+, n-1}$};
\node (mlr) at (5,3) [] {$\scriptstyle X^{(m)}_{-, n-1}$};
\node at (3,4) [] {$\cdots$};
\node (mll) at (3,3) [] {$\cdots$};
\node at (3,2) [] {$\cdots$};
\draw [->] (tr) -- (mrr);
\draw [->] (mrr) -- (br);
\draw [->] (tr) -- (mrl);
\draw [->] (mrl) -- (br);
\draw [->] (tm) -- (mmr);
\draw [->] (mmr) -- (bm);
\draw [->] (tm) -- (mml);
\draw [->] (mml) -- (bm);
\draw [->] (tl) -- (mlr);
\draw [->] (mlr) -- (bl);
\draw [->] (tl) -- (mll);
\draw [->] (mll) -- (bl);
\draw [->,dotted] (tr) -- (mmr);
\draw [->,dotted] (mmr) -- (br);
\draw [->,dotted] (tm) -- (mrl);
\draw [->,dotted] (mrl) -- (bm);
\draw [->,dotted] (tm) -- (mlr);
\draw [->,dotted] (mlr) -- (bm);
\draw [->,dotted] (tl) -- (mml);
\draw [->,dotted] (mml) -- (bl);
\end{tikzpicture}
\caption{The solid arrows to the right denote the action of $\tilde L_-$, those to the left of $\tilde L_+$. While the dotted arrows to the right indicate the action of $\tilde R_+$ and those to the left the one of $\tilde R_-$. }\label{fig:semtyp}
\end{figure}

\subsubsection{Atypical modules}

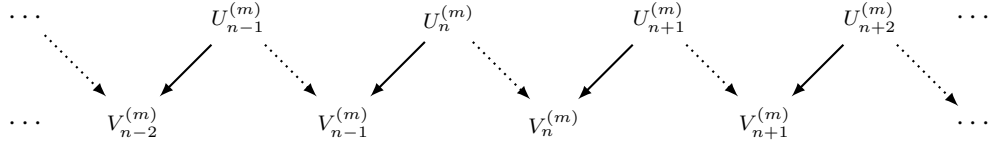
\begin{figure}[t]\centering
\begin{tikzpicture}[thick,>=latex,scale=1.4]
\node (t1) at (5,3) [] {$\scriptstyle U_n^{(m)}$};
\node (t2) at (7,3) [] {$\scriptstyle U_{n+1}^{(m)}$};
\node (t3) at (9,3) [] {$\scriptstyle U_{n+2}^{(m)}$};
\node (b1) at (4,2) [] {$\scriptstyle V_{n-1}^{(m)}$};
\node (b2) at (6,2) [] {$\scriptstyle V_n^{(m)}$};
\node (b3) at (8,2) [] {$\scriptstyle V_{n+1}^{(m)}$};
\node (t0) at (3,3) [] {$\scriptstyle U_{n-1}^{(m)}$};
\node (b0) at (2,2) [] {$\scriptstyle V_{n-2}^{(m)}$};
\node (t4) at (10,3) [] {$\cdots$};
\node (b4) at (10,2) [] {$\cdots$};
\node (tm) at (1,3) [] {$\cdots$};
\node (bm) at (1,2) [] {$\cdots$};
\draw [->] (t0) -- (b0);
\draw [->] (t1) -- (b1);
\draw [->] (t2) -- (b2);
\draw [->] (t3) -- (b3);
\draw [->,dotted] (tm) -- (b0);
\draw [->,dotted] (t0) -- (b1);
\draw [->,dotted] (t1) -- (b2);
\draw [->,dotted] (t2) -- (b3);
\draw [->,dotted] (t3) -- (b4);
\end{tikzpicture}
\caption{The solid arrows denote the action of $R_-$, the dotted ones the one of $L_-$. }\label{fig:atyp1}
\end{figure}

Let now $\tilde e=e=0$.
We will start by studying the left-right $\gl$ action.
We def\/ine
\begin{alignat*}{3}
&w_{(2, 2), n}  = \theta_-\tilde\theta_-\theta_+\tilde\theta_+ f_n, \qquad&&
w_{(2, -), n} = \tilde\theta_-\theta_+\tilde\theta_+ f_n, &\\
&w_{(2, +), n}  = \big(\theta_-\tilde\theta_-\tilde\theta_+ -i\tilde y \theta_-\tilde\theta_-\theta_+\big) f_n , \qquad&&
w_{(2, 0), n} = \big(\tilde\theta_-\tilde\theta_+ -i\tilde y \tilde\theta_-\theta_+\big) f_n , &\\
&w_{(-, 2), n}  = \big(\tilde\theta_-\theta_+\tilde\theta_+ -i\tilde y \theta_-\theta_+\tilde\theta_+\big) f_n, \qquad&&
w_{(-, -), n} = \tilde y\theta_+\tilde\theta_+ f_n, &\\
&w_{(-, +), n}  = \big(\tilde\theta_-\tilde\theta_+ -i\tilde y \big(\theta_-\tilde\theta_++\tilde \theta_-\theta_+\big) -\tilde y^2 \theta_-\theta_+\big) f_n , \qquad &&
w_{(-, 0), n} = \tilde y\big(\tilde\theta_+ -i\tilde y \theta_+\big) f_n , &\\
&w_{(+, 2), n}  = \theta_-\tilde\theta_-\tilde\theta_+ f_n, \qquad&&
w_{(+, -), n} = \tilde\theta_-\tilde\theta_+ f_n, & \\
&w_{(+, +), n}  = \tilde y \theta_-\tilde\theta_- f_n , \qquad&&
w_{(+, 0), n} = \tilde y \tilde\theta_- f_n ,& \\
&w_{(0, 2), n}  = \big(\tilde\theta_-\tilde\theta_+ -i\tilde y \theta_-\tilde\theta_+\big) f_n, \qquad&&
w_{(0, -), n} = \tilde y\tilde\theta_+ f_n,& \\
&w_{(0, +), n}  = \tilde y\big( \tilde \theta_--i\tilde y \theta_-\big) f_n , \qquad&&
w_{(0, 0), n} = \tilde y^2 f_n.&
\end{alignat*}
Further, def\/ine the one-dimensional vector spaces
\begin{gather*}
w_{(a, b), n}^{(m)}= \text{span}\big(\tilde y w_{(a, b), n}\big).
\end{gather*}
We compute the action of the left- and right-invariant vector f\/ields of the $\gl$ subalgebra. The result can be best visualized in the following diagram
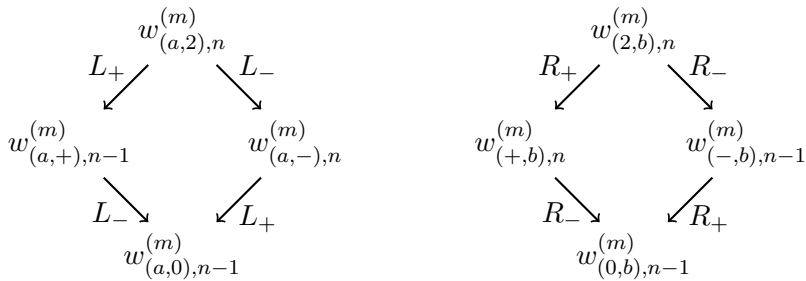
\begin{gather*}
\begin{tikzpicture}[auto,thick]
\node (top) at (0,1.5) [] {$w_{(a, 2), n}^{(m)}$};
\node (left) at (-1.5,0) [] {$w_{(a, +), n-1}^{(m)}$};
\node (right) at (1.5,0) [] {$w_{(a, -), n}^{(m)}$};
\node (bot) at (0,-1.5) [] {$w_{(a, 0), n-1}^{(m)}$};
\draw [->] (top) to (left);
\draw [->] (left) to (bot);
\draw [->] (top) to (right);
\draw [->] (right) to (bot);
\node at (-0.95,-0.95) [] {${L}_-$};
\node at (1,1) [] {${L}_-$};
\node at (-1,1) [] {${L}_+$};
\node at (1,-1) [] {${L}_+$};
\node (top2) at (6,1.5) [] {$w_{(2, b), n}^{(m)}$};
\node (left2) at (4.5,0) [] {$w_{(+, b), n}^{(m)}$};
\node (right2) at (7.5,0) [] {$w_{(-, b), n-1}^{(m)}$};
\node (bot2) at (6,-1.5) [] {$w_{(0, b), n-1}^{(m)}$};
\draw [->] (top2) to (left2);
\draw [->] (left2) to (bot2);
\draw [->] (top2) to (right2);
\draw [->] (right2) to (bot2);
\node at (5.05,-0.95) [] {$R_-$};
\node at (7,1) [] {$R_-$};
\node at (5,1) [] {$R_+$};
\node at (7,-1) [] {$R_+$};
\end{tikzpicture}
\end{gather*}
so that we observe
\begin{Proposition}
Under the left- and right-invariant vector fields of the $\gl$ subalgebra the vector space
\begin{gather*}
 B^{(m)}_n:= \bigoplus_{\substack{a\in \{2, +\} \\ b\in \{2, -\} }} w_{(a, b), n}^{(m)} \oplus
\bigoplus_{\substack{a\in \{2, +\} \\ b\in \{+, 0\} }} w_{(a, b), n-1}^{(m)} \oplus
\bigoplus_{\substack{a\in \{-, 0\} \\ b\in \{2, -\} }} w_{(a, b), n-1}^{(m)} \oplus
\bigoplus_{\substack{a\in \{-, 0\} \\ b\in \{+, 0\} }} w_{(a, b), n-2}^{(m)}
\end{gather*}
is the tensor product of $\gl$ projective covers,
\begin{gather*}
 B^{(m)}_n\cong\mathcal P_{2-n}^L \otimes \mathcal P_{n-2}^R.
 \end{gather*}
\end{Proposition}
We def\/ine
\begin{gather*}
{\mathcal P_n}^{(m)} := \bigoplus_{a\in\mathbb Z} B^{(m)}_{n+a}.
\end{gather*}
The second type of $\gl$ modules appearing are given by def\/ining the following
\begin{alignat*}{3}
& u_{2, n}^{(m)}  := \text{span}\big(\tilde y^m \theta_-\theta_+\tilde\theta_+ f_n  \big),\qquad&&
u_{R, n}^{(m)} := \text{span}\big(\tilde y^m \theta_-\tilde\theta_+ f_n  \big), &\\
& u_{L, n}^{(m)}  := \text{span}\big(\tilde y^m \big(\theta_-\tilde\theta_+-i\tilde y\theta_-\theta_+\big) f_n  \big),\qquad&&
u_{0, n}^{(m)} := \text{span}\big(\tilde y^m \tilde y\theta_- f_n  \big), &\\
& v_{2, n}^{(m)}  := \text{span}\big(\tilde y^m \theta_+\tilde\theta_+ f_n  \big),\qquad&&
v_{R, n}^{(m)} := \text{span}\big(\tilde y^m \tilde\theta_+ f_n  \big),& \\
& v_{L, n}^{(m)}  := \text{span}\big(\tilde y^m \big(\tilde\theta_+-i\tilde y\theta_+\big) f_n  \big),\qquad&&
v_{0, n}^{(m)} := \text{span}\big(\tilde y^m \tilde y f_n  \big).&
\end{alignat*}
Again, we compute the action of the left- and right-invariant vector f\/ields of the $\gl$ subalgebra. The result can be best visualized by diagrams. First, we restrict to the action of~$R_+$ and~$L_+$, we then get the modules for this subalgebra of~$\gl\oplus \gl$
\begin{gather*}
\begin{tikzpicture}[auto,thick]
\node (top) at (0,1.5) [] {$u_{ 2, n}^{(m)}$};
\node (left) at (-1.5,0) [] {$u_{R, n}^{(m)}$};
\node (right) at (1.5,0) [] {$u_{L, n-1}^{(m)}$};
\node (bot) at (0,-1.5) [] {$u_{0, n-1}^{(m)}$};
\draw [->] (top) to (left);
\draw [->] (left) to (bot);
\draw [->] (top) to (right);
\draw [->] (right) to (bot);
\node at (-0.95,-0.95) [] {${L}_+$};
\node at (1,1) [] {$L_+$};
\node at (-1,1) [] {${R}_+$};
\node at (1,-1) [] {${R}_+$};
\node (top2) at (6,1.5) [] {$v_{2, n}^{(m)}$};
\node (left2) at (4.5,0) [] {$v_{R, n}^{(m)}$};
\node (right2) at (7.5,0) [] {$v_{L, n-1}^{(m)}$};
\node (bot2) at (6,-1.5) [] {$v_{0, n-1}^{(m)}$};
\draw [->] (top2) to (left2);
\draw [->] (left2) to (bot2);
\draw [->] (top2) to (right2);
\draw [->] (right2) to (bot2);
\node at (5.05,-0.95) [] {$L_+$};
\node at (7,1) [] {$L_+$};
\node at (5,1) [] {$R_+$};
\node at (7,-1) [] {$R_+$};
\end{tikzpicture}
\end{gather*}
We call the f\/irst one $U_n^{(m)}$ and the second one $V_n^{(m)}$. The vector f\/ields~$L_-$ and~$R_-$ act trivially on the second one while the f\/irst one gives one-to-one maps illustrated in Fig.~\ref{fig:atyp1}. Thus the set
\begin{gather*}
{\mathcal P^-_n}^{(m)}:= \bigoplus_{a\in \mathbb Z} U_{n+a}^{(m)}\  \oplus\  \bigoplus_{b\in \mathbb Z} V_{n+b}^{(m)}
\end{gather*}
forms an inf\/inite-dimensional $\gl$ left-right module called~${\mathcal P^-_n}^{(m)}$.

\begin{figure}[t]\centering
\begin{tikzpicture}[thick,>=latex,scale=1.4]
\node (t1) at (5,3) [] {$\scriptstyle S_n^{(m)}$};
\node (t2) at (7,3) [] {$\scriptstyle S_{n+1}^{(m)}$};
\node (t3) at (9,3) [] {$\scriptstyle S_{n+2}^{(m)}$};
\node (b1) at (4,2) [] {$\scriptstyle T_{n-1}^{(m)}$};
\node (b2) at (6,2) [] {$\scriptstyle T_n^{(m)}$};
\node (b3) at (8,2) [] {$\scriptstyle T_{n+1}^{(m)}$};
\node (t0) at (3,3) [] {$\scriptstyle S_{n-1}^{(m)}$};
\node (b0) at (2,2) [] {$\scriptstyle T_{n-2}^{(m)}$};
\node (t4) at (10,3) [] {$\cdots$};
\node (b4) at (10,2) [] {$\cdots$};
\node (tm) at (1,3) [] {$\cdots$};
\node (bm) at (1,2) [] {$\cdots$};
\draw [->] (t0) -- (b0);
\draw [->] (t1) -- (b1);
\draw [->] (t2) -- (b2);
\draw [->] (t3) -- (b3);
\draw [->,dotted] (tm) -- (b0);
\draw [->,dotted] (t0) -- (b1);
\draw [->,dotted] (t1) -- (b2);
\draw [->,dotted] (t2) -- (b3);
\draw [->,dotted] (t3) -- (b4);
\end{tikzpicture}
\caption{The solid arrows denote the action of $L_+$, the dotted ones the one of $R_+$. }\label{fig:atyp2}
\end{figure}
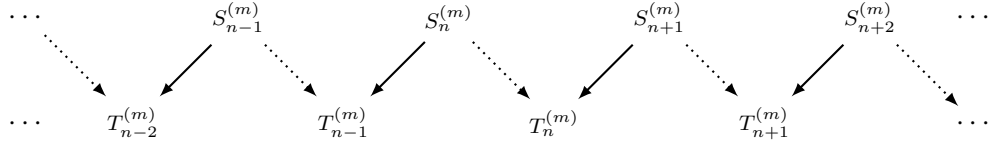

The third type of $\gl$ modules appearing is very similar to the previous one and are given by def\/ining the following
\begin{alignat*}{3}
& s_{2, n}^{(m)}  := \text{span}\big(\tilde y^m \theta_-\tilde\theta_-\theta_+ f_n  \big),\qquad&&
s_{L, n}^{(m)} := \text{span}\big(\tilde y^m \tilde\theta_-\theta_+ f_n  \big), & \\
& s_{R, n}^{(m)}  := \text{span}\big(\tilde y^m \big(\tilde\theta_-\theta_+-i\tilde y\theta_-\theta_+\big) f_n  \big),\qquad&&
s_{0, n}^{(m)} := \text{span}\big(\tilde y^m \tilde y\theta_+ f_n  \big), & \\
& t_{2, n}^{(m)}  := \text{span}\big(\tilde y^m \theta_-\tilde\theta_- f_n  \big),\qquad&&
t_{L, n}^{(m)} := \text{span}\big(\tilde y^m \tilde\theta_- f_n  \big),& \\
& t_{R, n}^{(m)}  := \text{span}\big(\tilde y^m \big(\tilde\theta_--i\tilde y\theta_-\big) f_n  \big),\qquad&&
t_{0, n}^{(m)} := \text{span}\big(\tilde y^m \tilde y f_n  \big).&
\end{alignat*}
Again, we compute the action of the left- and right-invariant vector f\/ields of the $\gl$ subalgebra. The result can also again be best visualized by diagrams. First, we restric to the action of~$R_-$ and~$L_-$, we then get the modules for this subalgebra of $\gl\oplus \gl$
\begin{gather*}
\begin{tikzpicture}[auto,thick]
\node (top) at (0,1.5) [] {$s_{ 2, n}^{(m)}$};
\node (left) at (-1.5,0) [] {$s_{R, n-1}^{(m)}$};
\node (right) at (1.5,0) [] {$s_{L, n}^{(m)}$};
\node (bot) at (0,-1.5) [] {$s_{0, n-1}^{(m)}$};
\draw [->] (top) to (left);
\draw [->] (left) to (bot);
\draw [->] (top) to (right);
\draw [->] (right) to (bot);
\node at (-0.95,-0.95) [] {${L}_-$};
\node at (1,1) [] {$L_-$};
\node at (-1,1) [] {${R}_-$};
\node at (1,-1) [] {${R}_-$};
\node (top2) at (6,1.5) [] {$t_{2, n}^{(m)}$};
\node (left2) at (4.5,0) [] {$t_{R, n-1}^{(m)}$};
\node (right2) at (7.5,0) [] {$t_{L, n}^{(m)}$};
\node (bot2) at (6,-1.5) [] {$t_{0, n-1}^{(m)}$};
\draw [->] (top2) to (left2);
\draw [->] (left2) to (bot2);
\draw [->] (top2) to (right2);
\draw [->] (right2) to (bot2);
\node at (5.05,-0.95) [] {$L_-$};
\node at (7,1) [] {$L_-$};
\node at (5,1) [] {$R_-$};
\node at (7,-1) [] {$R_-$};
\end{tikzpicture}
\end{gather*}
We call the f\/irst one $S_n^{(m)}$ and the second one $T_n^{(m)}$. The vector f\/ields $L_+$ and $R_+$ act trivially on the second one while the f\/irst one gives one-to-one maps illustrated in Fig.~\ref{fig:atyp2}. Thus the set
\begin{gather*}
{\mathcal P^+_n}^{(m)}:= \bigoplus_{a\in \mathbb Z} S_{n+a}^{(m)}\  \oplus\  \bigoplus_{b\in \mathbb Z} T_{n+b}^{(m)}
\end{gather*}
forms an inf\/inite-dimensional $\gl$ left-right module called ${\mathcal P^+_n}^{(m)}$.

The forth type of $\gl$ module is given by
\begin{alignat*}{3}
& r_{2, n}^{(m)}  := \text{span}\big(\tilde y^m \theta_-\theta_+ f_n  \big),\qquad&&
r_{-, n}^{(m)} := \text{span}\big(\tilde y^m \theta_+ f_n  \big), &\\
& r_{+, n}^{(m)}  := \text{span}\big(\tilde y^m \theta_- f_n  \big),\qquad&&
r_{0, n}^{(m)} := \text{span}\big(\tilde y^m f_n  \big).&
\end{alignat*}
Then as a left-right module we get the picture as in Fig.~\ref{fig:atyp3}. We denote this module by~${\mathcal P^0_n}^{(m)}$.
\begin{figure}[t]\centering
\begin{tikzpicture}[thick,>=latex,scale=1.4]
\node (tr) at (12,4) [] {$\scriptstyle r^{(m)}_{2, n+1}$};
\node (tm) at (8,4) [] {$\scriptstyle r^{(m)}_{2, n}$};
\node (tl) at (4,4) [] {$\scriptstyle r^{(m)}_{2, n-1}$};
\node (br) at (12,2) [] {$\scriptstyle r^{(m)}_{0, n}$};
\node (bm) at (8,2) [] {$\scriptstyle r^{(m)}_{0, n-1}$};
\node (bl) at (4,2) [] {$\scriptstyle r^{(m)}_{0, n-2}$};
\node at (13,4) [] {$\cdots$};
\node (mrr) at (13,3) [] {$\cdots$};
\node at (13,2) [] {$\cdots$};
\node (mrl) at (11,3) [] {$\scriptstyle r^{(m)}_{+, n}$};
\node (mmr) at (9,3) [] {$\scriptstyle r^{(m)}_{-, n}$};
\node (mml) at (7,3) [] {$\scriptstyle r^{(m)}_{+, n-1}$};
\node (mlr) at (5,3) [] {$\scriptstyle r^{(m)}_{-, n-1}$};
\node at (3,4) [] {$\cdots$};
\node (mll) at (3,3) [] {$\cdots$};
\node at (3,2) [] {$\cdots$};
\draw [->] (tr) -- (mrr);
\draw [->] (mrr) -- (br);
\draw [->] (tr) -- (mrl);
\draw [->] (mrl) -- (br);
\draw [->] (tm) -- (mmr);
\draw [->] (mmr) -- (bm);
\draw [->] (tm) -- (mml);
\draw [->] (mml) -- (bm);
\draw [->] (tl) -- (mlr);
\draw [->] (mlr) -- (bl);
\draw [->] (tl) -- (mll);
\draw [->] (mll) -- (bl);
\draw [->,dotted] (tr) -- (mmr);
\draw [->,dotted] (mmr) -- (br);
\draw [->,dotted] (tm) -- (mrl);
\draw [->,dotted] (mrl) -- (bm);
\draw [->,dotted] (tm) -- (mlr);
\draw [->,dotted] (mlr) -- (bm);
\draw [->,dotted] (tl) -- (mml);
\draw [->,dotted] (mml) -- (bl);
\end{tikzpicture}
\caption{The solid arrows to the right denote the action of~$L_-$, those to the left of~$L_+$. While the dotted arrows to the right indicate the action of $R_+$ and those to the left the one of $R_-$. }\label{fig:atyp3}
\end{figure}
Observe that we have the inclusion of $\gl$ left-right modules
\begin{gather*}
{\mathcal P^0_n}^{(m)} \subset {\mathcal P^+_n}^{(m-1)} \oplus {\mathcal P^-_n}^{(m-1)}, \qquad
{\mathcal P^\pm_n}^{(m)} \subset {\mathcal P_n}^{(m-1)}
\end{gather*}
for $m>0$. Especially as a $\gl$ left-right module the atypical part, that is the $e=\tilde e=0$ part of the space of functions on $\tgl$ decomposes as
\begin{gather*}
\int_0^1 \left( {\mathcal P^0_n}^{(0)} \oplus {\mathcal P^1_n}^{(0)}  \oplus \bigoplus_{m=1}^\infty {\mathcal P_n}^{(m)}\right) dn.
\end{gather*}
Here ${\mathcal P^1_n}^{(0)}$ denotes the module generated by ${\mathcal P^+_n}^{(0)}$ and  ${\mathcal P^-_n}^{(0)}$ whose intersection is the submodule~${\mathcal P^0}^{(1)}$.
These modules are constructed such that the tilded action is easy to read of, the~$\tilde L_\pm$ and~$\tilde R_\pm$ are $\SLSA{sl}{1}{1}$ module homomorphism (they only change the action of~$L_N$ or~$R_N$) and for $m>0$ they act as
\begin{gather*}
\text{Image}\big( \tilde L_\pm\colon \, {\mathcal P_n}^{(m)} \longrightarrow {\mathcal P_n}^{(m-1)}\big)  =
\text{Image}\big( \tilde R_\pm\colon \,  {\mathcal P_n}^{(m)} \longrightarrow {\mathcal P_n}^{(m-1)}\big) = {\mathcal P_n^\pm}^{(m)}, \\
\text{Image}\big( \tilde L_\pm\colon \, {\mathcal P_n^\mp}^{(m)} \longrightarrow {\mathcal P_n^\mp}^{(m-1)}\big)  =
\text{Image}\big( \tilde R_\pm\colon \, {\mathcal P_n^\mp}^{(m)} \longrightarrow {\mathcal P_n^\mp}^{(m-1)}\big) = {\mathcal P_n^0}^{(m)},
\end{gather*}
while in the $m=0$ case we have
\begin{gather*}
\text{Image}\big( \tilde L_\pm\colon \, {\mathcal P_n}^{(0)} \longrightarrow {\mathcal P_n^\pm}^{(0)}\big)  =
\text{Image}\big( \tilde R_\pm\colon \, {\mathcal P_n}^{(0)} \longrightarrow {\mathcal P_n^\pm}^{(0)}\big) = {\mathcal P_n^\pm}^{(0)}, \\
\text{Image}\big( \tilde L_\pm\colon \, {\mathcal P_n^\mp}^{(0)} \longrightarrow {\mathcal P_n^0}^{(0)}\big)  =
\text{Image}\big( \tilde R_\pm\colon \, {\mathcal P_n^\mp}^{(0)} \longrightarrow {\mathcal P_n}^{(0)}\big) = {\mathcal P_n^0}^{(0)}.
\end{gather*}
The image of these maps is in the kernel and we inspect that it is precisely the kernel, hence both $\tilde L_+$ and $\tilde R_+$ action is described by the long-exact sequences
\begin{gather*}
 \cdots \longrightarrow {\mathcal P_n}^{(m)}  \longrightarrow {\mathcal P_n}^{(m-1)}  \longrightarrow \dots\longrightarrow {\mathcal P_n}^{(1)}  \longrightarrow {\mathcal P_n}^{(0)}
\longrightarrow {\mathcal P_n^+}^{(0)}  \longrightarrow 0, \\
 \cdots \longrightarrow {\mathcal P_n^-}^{(m)}  \longrightarrow {\mathcal P_n^-}^{(m-1)}  \longrightarrow \dots\longrightarrow {\mathcal P_n^-}^{(1)}  \longrightarrow {\mathcal P_n^-}^{(0)}
\longrightarrow {\mathcal P_n^0}^{(0)}  \longrightarrow 0,
\end{gather*}
while the one of  $\tilde L_-$ and $\tilde R_-$ action is described by the long-exact sequences
\begin{gather*}
 \cdots \longrightarrow {\mathcal P_n}^{(m)}  \longrightarrow {\mathcal P_n}^{(m-1)}  \longrightarrow \dots\longrightarrow {\mathcal P_n}^{(1)}  \longrightarrow {\mathcal P_n}^{(0)}
\longrightarrow {\mathcal P_n^-}^{(0)}  \longrightarrow 0, \\
 \cdots \longrightarrow {\mathcal P_n^+}^{(m)}  \longrightarrow {\mathcal P_n^+}^{(m-1)}  \longrightarrow \dots\longrightarrow {\mathcal P_n^+}^{(1)}  \longrightarrow {\mathcal P_n^+}^{(0)}
\longrightarrow {\mathcal P_n^0}^{(0)}  \longrightarrow 0.
\end{gather*}
The action of $\tilde L_N$ and $\tilde R_N$ is uniquely specif\/ied by its action on the top components, the map can be split into two components $\tilde L_N=\tilde L_N^a +\tilde L_N^b$ and
$\tilde R_N=\tilde R_N^a +\tilde R_N^b$ such that~$\tilde L_N^a$ and~$\tilde R_N^a$ are both one-to-one maps from~$x_{i}^{(m)}$ to~$x_i^{(m-1)}$ for all $x$ in $\{s, t, u, v, w\}$ and $i$ labelling the corresponding indices.  More precisely they act as multiplication by~$\pm im$ ($+$ for~$\tilde R$ and $-$ for $\tilde L$) times~$\tilde y^{-1}$. Especially they act trivially when $m=0$.
The images of the second part are (for~$m\ne 0$)
\begin{gather*}
\text{Image}\big( \tilde L_N^b\colon \, {\mathcal P_n}^{(m)} \longrightarrow {\mathcal P_n}^{(m-1)}\big)  =
\text{Image}\big( \tilde R_N^b\colon \, {\mathcal P_n}^{(m)} \longrightarrow {\mathcal P_n}^{(m-1)}\big) = {\mathcal P_n^1}^{(m)}, \\
\text{Image}\big( \tilde L_N^b\colon \, {\mathcal P_n^\pm}^{(m)} \longrightarrow {\mathcal P_n^\pm}^{(m-1)}\big)  =
\text{Image}\big( \tilde R_N^b\colon \, {\mathcal P_n^\pm}^{(m)} \longrightarrow {\mathcal P_n^\pm}^{(m-1)}\big) = {\mathcal P_n^0}^{(m)}, \\
\text{Image}\big( \tilde L_N^b\colon \, {\mathcal P_n^0}^{(m)} \longrightarrow {\mathcal P_n^0}^{(m-1)}\big)  =
\text{Image}\big( \tilde R_N^b\colon \, {\mathcal P_n^0}^{(m)} \longrightarrow {\mathcal P_n^0}^{(m-1)}\big) =0 ,
\end{gather*}
and for $m=0$
\begin{gather*}
\text{Image}\big( \tilde L_N^b\colon \, {\mathcal P_n}^{(0)} \longrightarrow {\mathcal P_n^1}^{(0)}\big)  =
\text{Image}\big( \tilde R_N^b\colon \, {\mathcal P_n}^{(0)} \longrightarrow {\mathcal P_n^1}^{(0)}\big) = {\mathcal P_n^1}^{(0)}, \\
\text{Image}\big( \tilde L_N^b\colon \, {\mathcal P_n^\pm}^{(0)} \longrightarrow {\mathcal P_n^0}^{(0)}\big)  =
\text{Image}\big( \tilde R_N^b\colon \, {\mathcal P_n^\pm}^{(0)} \longrightarrow {\mathcal P_n^0}^{(0)}\big) = {\mathcal P_n^0}^{(0)}, \\
\text{Image}\big( \tilde L_N^b\colon \, {\mathcal P_n^0}^{(0)} \longrightarrow {\mathcal P_n^0}^{(0)}\big)  =
\text{Image}\big( \tilde R_N^b\colon \, {\mathcal P_n^0}^{(0)} \longrightarrow {\mathcal P_n^0}^{(0)}\big) = 0.
\end{gather*}

\section{Free f\/ield realization}\label{section3}

In this section we construct free f\/ield realization of the holomorphic and anti-holomorphic currents of af\/f\/ine Takif\/f super Lie algebra~$\tgla$, and its modules.
Such free f\/ield realizations are constructed by starting with a triangular decomposition of a Lie supergroup element as in~\eqref{eq:triangular}. This procedure is outlined in~\cite{QS}, and we can immediately infer the bulk screening charge from~(3.2.32) of~\cite{C1}, as well as the currents in the free f\/ield realization from~(3.2.39) of that work. We however still need to prove that the screening charge has the desired properties. The harmonic analysis is expected to be a semi-classical (large level) limit of the screened free f\/ield theory. Examples on this for type one Lie supergroups are~\cite{CQS,QS, SS2,SS1}.

The Takif\/f algebra admits an automorphism $\omega_\alpha$ for $\alpha$ in $\mathbb{C}$.
It is given by
\begin{gather*}
E\mapsto E + \alpha \tilde E, \qquad  \psi^+\mapsto \psi^++\alpha \tilde\psi^+
\end{gather*}
and leaving all others invariant.
It allows us to restrict to the bilinear form $\tilde\kappa$. In order to see this, consider the bilinear form $\kappa_\beta:=\tilde \kappa +\beta\kappa_0$, then
\begin{gather*}
\kappa( \omega_{-\beta}(X), \omega_{-\beta}(Y)) =\tilde\kappa(X, Y)
\end{gather*}
for all $X$, $Y$ in $\tgl$.
In~\cite{BR}, the af\/f\/inization of $\tgl$ was discussed and levels $\tilde k$ and $k$ associated to the bilinear forms~$\tilde \kappa$ and~$\kappa$ were introduced.
The above automorphism relates the af\/f\/ine vertex algebra of~$\tgl$ at levels $(\tilde k, k)$ to the one at levels~$(\tilde k, 0)$.

\subsection{Currents and screening operator}
Consider a holomorphic $bc$-ghost vertex algebra taking values in the odd part of the Takif\/f superalgebra. Its components have OPE
\begin{gather*}
b(z) \tilde c(w) \sim \tilde b(z) c(w) \sim \frac{1}{(z-w)}
\end{gather*}
and all other OPEs are regular. We also need free bosons (generators of the Heisenberg VOA) taking values in the even part of the Takif\/f superalgebra. Its components have OPE in their holomorphic part
\begin{gather*}
\partial X(z) \partial\tilde Y(w) \sim \partial\tilde X(z) \partial Y(w) \sim \frac{-1/k}{(z-w)^2}.
\end{gather*}
In the same way we def\/ine anti holomorphic partners of the ghosts. The f\/ields $X(z, \bar z)$, $Y(z, \bar z)$, $\tilde X(z, \bar z)$, $\tilde Y(z, \bar z)$  will be considered as both holomorphic and anti holomorphic, i.e., depending on~$z$ and~$\bar{z}$, with the same OPEs in anti-holomorphic sector.

Currents should be thought of as quantizations of invariant vector f\/ields. We def\/ine holomorphic currents by the formulas
%outlined in my thesis (carefull elements are allways identif\/ied using $\tilde \kappa$), that is
\begin{alignat}{3}
& J^N  = -i\partial \tilde X +b\tilde c+ \tilde b c+i\partial Y, \qquad&& J^E= -ik\partial \tilde Y,& \nonumber\\
& J^+  = \tilde b,\qquad&&  J^-= k\partial \tilde c -ik\tilde c \partial Y -ikc\partial \tilde Y, &\nonumber \\
& \tilde J^N  = -i\partial X +bc, \qquad&& \tilde J^E=-ik\partial Y, & \nonumber \\
& \tilde J^+  = b,\qquad && \tilde J^-= k\partial c -ikc \partial Y.&\label{holcur}
\end{alignat}
and the anti holomorphic ones by
\begin{alignat}{3}
& \bar J^N  = ik\bar{\partial}\tilde{X}-i\bar{\partial
}Y-\bar{b}\bar{\tilde{c}}-\bar{\tilde{b}}\bar{c}, \qquad && \bar J^E=ik\bar \partial \tilde Y, & \nonumber\\
& \bar J^+  = k\bar\partial \tilde{ \bar c} -ik\tilde{ \bar c} \bar \partial Y -ik\bar c\bar \partial \tilde Y ,\qquad&&  \bar J^-= -\bar{\tilde b}, & \nonumber\\
& \tilde{\bar J}^N  = ik\bar\partial X -\bar b\bar c, \qquad&&  \tilde{ \bar J}^E=ik\bar\partial Y, & \nonumber \\
& \tilde{\bar J}^+  = k\bar\partial \bar c -ik\bar c \bar \partial Y,\qquad&&  \tilde{\bar J}^-=-\bar b .&\label{aholcur}
\end{alignat}
By direct calculation one can see that

\begin{Proposition}\label{prop:cural}
The OPEs of currents \eqref{holcur} are those of the level $k$ with respect to $\tilde \kappa$ affine Takiff superalgebra, that is
the only non-regular OPEs are
\begin{alignat*}{3}
& J^N(z)J^\pm(w)  \sim \pm\frac{J^\pm(w)}{(z-w)}, \qquad&&  J^+(z)J^-(w)\sim \frac{J^E(w)}{(z-w)},& \\
& J^N(z) \tilde J^E(w)  \sim \frac{k}{(z-w)^2}, \qquad&& J^N(z) \tilde J^\pm(w) \sim \pm\frac{\tilde J^\pm(w)}{(z-w)},& \\
& \tilde J^N(z) J^E(w)  \sim \frac{k}{(z-w)^2}, \qquad&& \tilde J^N(z)  J^\pm(w) \sim \pm\frac{\tilde J^\pm(w)}{(z-w)},&\\
& J^+(z) \tilde J^-(w)  \sim \frac{k}{(z-w)^2}+ \frac{\tilde J^E(w)}{(z-w)}, \qquad && J^-(z) \tilde J^+(w) \sim -\frac{k}{(z-w)^2}+ \frac{\tilde J^E(w)}{(z-w)}.&
\end{alignat*}
and the same set of OPEs of anti-holomorphic currents~\eqref{aholcur}.
\end{Proposition}

One can substitute these currents into the energy-momentum tensor obtained in~\cite{BR} (see also~\cite{Mo})
\begin{gather*}
T(z)=\frac{1}{k} \, {\colon}\!\!\left(J^N\tilde{J}^E + J^E\tilde{J}^N - J^+\tilde{J}^- + J^-\tilde{J}^++\frac{1}{k}\tilde{J}^E\tilde{J}^E\right)\!{\colon} (z)
\end{gather*}
getting
\begin{gather}
T(z)={\colon}\!\big(\partial\tilde{X}\partial Y+\partial\tilde{Y}\partial X-k\partial^2 Y-\tilde{b}\partial c -b\partial\tilde{c})\big){\colon}(z), \label{T}
\end{gather}
which clearly has central charge zero.
The key role in free f\/ield realization is played by screening operators.
\begin{Theorem}\label{Theor:scrin}
The currents~\eqref{holcur} lie in the joint kernel of the intertwiners
\begin{gather*}
S= \oint b(z) e^{iY(z)}dz, \qquad  \tilde S= \oint \big(i\tilde Y(z) b(z)+ \tilde b(z)\big) e^{iY(z)}dz.
\end{gather*}
\end{Theorem}

\begin{proof}
The computations are straightforward. We demonstrate here only action of~$S$ and~$\tilde S$ on~$J^-$ and~$\tilde J^-$. For~$S$ we get
\begin{gather*}
S J^-(w) = \oint ke^{iY(z)} \left( \frac{1}{(z-w)^2} +\frac{-i\partial Y(w)}{(z-w)}+\text{regular}\right) dz \\
\hphantom{S J^-(w)}{}
= \oint \left(e^{iY(w)}\frac{k}{(z-w)^2}+\text{regular}\right) dz=0,\\
S \tilde J^-(w) = \oint \text{regular}\  dz=0,
\end{gather*}
and for the action of $\tilde S$ it becomes
\begin{gather*}
\tilde S J^-(w)  =  \oint e^{iY(z)} \left( \frac{ik\tilde Y(z)}{(z-w)^2} -\frac{ik\partial \tilde Y(w) - k\tilde Y(w)\partial Y(w)}{(z-w)}+\text{regular}\right) dz \\
\hphantom{\tilde S J^-(w)}{}
=  \oint \left( e^{iY(w)} \frac{ik\tilde Y(w)}{(z-w)^2} +\text{regular}\right) dz =0,\\
\tilde S \tilde J^-(w)   = \oint e^{iY(z)} \left( \frac{k}{(z-w)^2} -\frac{ik\partial Y(w)}{(z-w)}+\text{regular}\right) dz \\
\hphantom{\tilde S \tilde J^-(w)}{}
 = \oint \left(e^{iY(w)}\frac{k}{(z-w)^2}+\text{regular}\right) dz=0.\tag*{\qed}
\end{gather*}
\renewcommand{\qed}{}
\end{proof}

The same theorem can be proved for the anti-holomorphic part.
 We can now def\/ine bulk screening charge which involves both holomorphic and anti-holomorphic parts:
\begin{gather*}
Q = \int  \big(\bar b(\bar z) \tilde b(z) + \tilde{\bar{b}}(\bar z) b(z) +i \bar b(\bar z) b(z) \tilde Y(z, \bar z) \big) e^{iY(z, \bar z)} d^2z.
\end{gather*}
This $Q$ should def\/ine correlation functions, which will not be considered in this paper.

\subsection{Vertex operators}

We def\/ine the vertex operator
\begin{gather*}
V_{e, n, \tilde e, \tilde n}= {\colon}\! e^{ieX-inY+i\tilde e \tilde X- i\tilde n \tilde Y}{\colon}\!,
\end{gather*}
whose conformal dimension can be calculated with (\ref{T}):
\begin{gather}\label{cd}
\Delta= e\tilde n +n\tilde e + i k\tilde e.
\end{gather}

It transforms as a primary f\/ield corresponding to a highest-weight state, that is
\begin{gather*}
J^E(z)V_{e, n, \tilde e, \tilde n}(w)  \sim \frac{keV_{e, n, \tilde e, \tilde n}(w)}{(z-w)}, \\
J^N(z)V_{e, n, \tilde e, \tilde n}(w)   \sim \frac{(n-\tilde e)V_{e, n, \tilde e, \tilde n}(w)}{(z-w)}, \\
\tilde J^E(z)V_{e, n, \tilde e, \tilde n}(w)  \sim \frac{k\tilde eV_{e, n, \tilde e, \tilde n}(w)}{(z-w)}, \\
\tilde J^N(z)V_{e, n, \tilde e, \tilde n}(w)  \sim \frac{\tilde nV_{e, n, \tilde e, \tilde n}(w)}{(z-w)}, \\
J^-(z)V_{e, n, \tilde e, \tilde n}(w)  \sim \frac{k e\, {\colon}\!c(w)V_{e, n, \tilde e, \tilde n}(w){\colon}+\tilde e k \, {\colon}\!\tilde c(w) V_{e, n, \tilde e, \tilde n}(w){\colon}}{(z-w)}, \\
\tilde J^-(z)V_{e, n, \tilde e, \tilde n}(w)  \sim \frac{k\tilde e\, {\colon}\!c(w)V_{e, n, \tilde e, \tilde n}(w){\colon}}{(z-w)}, \\
J^-(z){\colon}c(w)V_{e, n, \tilde e, \tilde n}(w){\colon}  \sim \frac{k\tilde e\, {\colon}\!\tilde c(w) c(w)V_{e, n, \tilde e, \tilde n}(w){\colon}}{(z-w)}.
\end{gather*}

\subsection{Free f\/ield realization of typical modules}

We can now combine holomorphic and anti-holomorphic ghost f\/ields with the vertex operator introduced above to form bulk free f\/ield realization for typical modules. It is convenient to represent it in the matrix form
\begin{gather*}
\phi_{e, n, \tilde e, \tilde n}(z)= {\colon}\!\tilde{Y}^mV_{e, n, \tilde e, \tilde n}{\colon}\!(z,\bar z) \begin{pmatrix}  1 & \bar c & \tilde{\bar c} & \tilde{\bar c}\bar c \\
c &  c\bar c & c\tilde{\bar c} & c\tilde{\bar c}\bar c  \\
\tilde c & \tilde c\bar c & \tilde c\tilde{\bar c} & \tilde c\tilde{\bar c}\bar c \\
c\tilde c & c\tilde c\bar c & c\tilde c\tilde{\bar c} & c\tilde c\tilde{\bar c}\bar c
\end{pmatrix}\!{\colon}\!.
\end{gather*}
Here each row and column of the matrix $C$ is understood as a span of its elements. Using the OPEs one can easily check that columns carry typical representations for the holomorphic currents and the rows~-- for the anti-holomorphic ones. The elements~$C_{1,i}$ are the highest weights for the columns, and the elements~$C_{i,1}$~-- for the rows. For $m=0$ they are Virasoro primary f\/ields with conformal dimension~(\ref{cd}).

\subsection{A free f\/ield realization for atypical and semitypical modules}

Consider two pairs of symplectic fermions $\chi_\pm$, $\tilde\chi_\pm$ with OPE
\begin{gather*}
\chi_+(z)\tilde\chi_-(w) \sim \chi_-(z)\tilde\chi_+(w) \sim \frac{k}{(z-w)^2}
\end{gather*}
and a pair of pairs of free bosons $\partial X$, $\partial Y$, $\partial \tilde X$, $\partial \tilde Y$ with OPE
\begin{gather*}
\partial X(z)\partial \tilde Y(w) \sim \partial Y(z)\partial \tilde X(w) \sim \frac{1}{(z-w)^2}.
\end{gather*}
Recall, that symplectic fermions are a logarithmic conformal f\/ield theory. Especially, the vacuum is part of a larger indecomposable module whose four composition factors are all isomorphic to the vacuum itself, for details see~\cite{K}. The generating f\/ields of these modules are $\Omega_\pm(z, \bar z)$, $\tilde\theta_\pm(z, \bar z)$, $\theta_\pm(z, \bar z)$ with the following behaviour under the action of the symplectic fermions:
\begin{alignat*}{3}
& \chi_\pm(z)\Omega_\pm(w, \bar w)  \sim \frac{k\theta_\pm(w, \bar w)}{(z-w)}, \qquad &&
\tilde\chi_\mp(z)\Omega_\pm(w, \bar w) \sim \frac{k\tilde\theta_\mp(w, \bar w)}{(z-w)},& \\
& \chi_\pm(z)\tilde\theta_\mp(w, \bar w)  \sim \frac{k}{(z-w)}, \qquad && \tilde\chi_\mp(z)\theta_\pm(w, \bar w) \sim -\frac{k}{(z-w)}.&
\end{alignat*}
We split the f\/ield $\tilde Y(z, \bar z)$ into chiral and anti-chiral part, that is $\tilde Y(z, \bar z)=\tilde Y(z)+\tilde{\bar Y}(\bar z)$.
Then we def\/ine
\begin{gather}
K^E(z) = k\partial \tilde Y(z), \qquad
K^N(z) = \partial \tilde X(z), \qquad
\tilde K^E(z) = k\partial Y(z), \qquad
\tilde K^N(z) = \partial X(z), \nonumber\\
K^\pm(z)  = e^{\pm Y(z)} \big( \chi_\pm(z) -\tilde Y(z) \tilde\chi_\pm(z) \big), \qquad
\tilde K^\pm(z) = \mp e^{\pm Y(z)} \tilde\chi_\pm(z).\label{eq:SFcur}
\end{gather}
and in analogy also anti-holomorphic currents.
We compute
\begin{Proposition}\label{prop:curalSF}
The OPEs of currents~\eqref{eq:SFcur} are those of the level~$k$ with respect to~$\tilde \kappa$ affine Takiff superalgebra.
These OPEs are listed in Proposition~{\rm \ref{prop:cural}}.
The energy-momentum tensor for these currents is
\begin{gather*}
T(z)  = {\colon}\!\partial \tilde X(z) \partial Y(z){\colon}\! +{\colon}\!\partial \tilde Y(z) \partial X(z){\colon}\!-\frac{1}{k} \big({\colon}\!\chi_+(z)\tilde\chi_-(z){\colon}\!+{\colon}\!\chi_-(z)\tilde\chi_+(z){\colon}\!\big).
\end{gather*}
\end{Proposition}

These currents are in the kernel of
\begin{gather*}
Q:= \oint \big({\colon}\!\tilde\chi_+(z)\tilde\theta_+(z, \bar z){\colon}\! +{\colon}\!\tilde\chi_-(z)\tilde\theta_-(z, \bar z){\colon}\!+2k\partial X(z)\big)dz
\end{gather*}
acting on the chiral algebra generated under operator product by $e^{\pm Y(z)}$, $\tilde Y(z)$, $\chi_\pm(z)$, $\tilde\chi_\pm(z)$, $\partial X(z)$, $\partial \tilde X(z)$ and their derivatives. The currents are also invariant under the $U(1)$-action induced by $\varphi$ which is def\/ined by
\begin{gather*}
\varphi\big(e^{\pm Y(z)}\big) = \mp e^{\pm Y(z)}, \qquad \varphi (\chi_\pm(z) ) = \pm\chi_\pm(z), \qquad
 \varphi (\tilde\chi_\pm(z) ) = \pm\tilde\chi_\pm(z).
\end{gather*}
As discussed in~\cite{BR}, semi-typical and atypical modules appear when $\tilde e/k=m\in {\mathbb Z}$.
The vertex operator
\begin{gather*}
V_{e, n, \tilde e, \tilde n}(z) := {\colon}\!e^{\frac{e}{k}X(z)+nY(z)+\frac{\tilde e}{k}\tilde X(z)+\tilde n\tilde Y(z)}{\colon}
\end{gather*}
has weight $(e, n, \tilde e, \tilde n)$ under the currents $K^E$, $K^N$, $\tilde K^E$, $\tilde K^N$.
Def\/ine
\begin{gather*}
\phi^\pm_m(z) := {\colon}\!\chi_\pm(z)\partial\chi_\pm(z) \cdots \partial^m\chi_\pm(z){\colon}\!, \qquad
\tilde\phi^\pm_m(z) := {\colon}\!\tilde\chi_\pm(z)\partial\tilde\chi_\pm(z) \cdots \partial^m\tilde\chi_\pm(z){\colon}\!.
\end{gather*}
Then it is a computation to verify that for $m\leq 0$
\begin{gather*}
\varphi_{e, n, mk, \tilde n}(z) :=V_{e, n, mk, \tilde n}(z) \phi^+_{-m}(z)\tilde\phi^+_{-m}
\end{gather*}
is a primary f\/ield for the highest-weight representation of weight $(e, n, mk, \tilde n)$ of the correct conformal dimension $e\tilde n/k +mn+m(m-1)$.
It is a submodule of the modules generated by
\begin{gather*}
\varphi_{e, n, mk, \tilde n}(z)\theta_+(z, \bar z), \qquad
\varphi_{e, n, mk, \tilde n}(z)\tilde\theta_+(z, \bar z), \qquad
\varphi_{e, n, mk, \tilde n}(z)\theta_+(z, \bar z)\tilde\theta_+(z, \bar z).
\end{gather*}
If $m\geq0$, then we can construct analogous primary f\/ields for lowest-weight representations by replacing $\pm$, i.e., the primary f\/ield is
\begin{gather*}
\varphi_{e, n, mk, \tilde n}(z):=V_{e, n, mk, \tilde n}(z) \phi^-_{m}(z)\tilde\phi^-_{m}
\end{gather*}
and it is a submodule of the modules generated by
\begin{gather*}
\varphi_{e, n, mk, \tilde n}(z)\theta_-(z, \bar z), \qquad
\varphi_{e, n, mk, \tilde n}(z)\tilde\theta_-(z, \bar z), \qquad
\varphi_{e, n, mk, \tilde n}(z)\theta_-(z, \bar z)\tilde\theta_-(z, \bar z).
\end{gather*}
So in all these cases, we can use our free f\/ield realization to construct indecomposable but reducible modules.
In the case of $m=0$, we have both highest and lowest weight modules. These modules now deserve further investigation.
We plan to study the combined left-right action of currents on these modules in the near future. The results then have to be compared to the semi-typicals of our harmonic analysis.
Finally, let us note that there is a third free f\/ield realization simply by replacing~$\tilde Y(z)$ and~$\partial X(z)$ by a~$\beta\gamma$ bosonic ghost algebra. This one will also be investigated later.

\section{Conclusion}\label{section4}

This work together with~\cite{BR} puts one in the situation to study conformal f\/ield theories with chiral algebra the af\/f\/inization of $\tgl$.
We have studied the harmonic analysis on the supergroup in detail and especially we have found that all irreducible modules are part of larger indecomposable but reducible modules.
We then have found two free f\/ield realizations for the associated CFT. The next question is to understand the spectrum of the full CFT better. For this, one needs to study the free f\/ield modules in detail. Especially the structure of semi-typicals and atypicals from the symplectic fermion free f\/ield realization needs to be worked out. In logarithmic CFT, indecomposable structure of modules is related to logarithmic singularities in correlation functions. Screening charges provide integral formula for correlation functions, so that one can use our f\/indings to compute those. Since all our modules in the harmonic analysis are submodules of indecomposable but reducible ones we expect that logarithmic singularities will be a generic feature. This will be in strong contrast to logarithmic CFTs based on non Takif\/f supergroups.

Our f\/indings have the drawback, that the Takif\/f superalgebra forced us to allow for polynomials in the variable $\tilde y$. In the conformal f\/ield theory this suggests to also allow for polynomials in the f\/ield $\tilde y$.
This is unusual, as, e.g., the free boson CFT does not allow for f\/ields of analogous type. Of course the OPE involving such polynomials will automatically yield logarithmic singularities. As mentioned in the end of last section, a possible way out is to replace~$\tilde Y$ and~$\partial Y$ by a~$\beta\gamma$ bosonic ghost algebra. In that case we would have no semi-typical modules but still typical and atypical ones.

There are many future directions as boundary states and boundary correlation functions. These can surely be studied along the lines of their non-Takif\/f analouges~\cite{CQS, CS}. Recall, that in that case also harmonic analysis was the important starting point. We are more interested in Takif\/f superalgebra CFTs based on more complicated algebras. It was shown that $\AKMSA{gl}{1}{1}$ has many interesting simple current extensions~\cite{CR1, CR2}, including $\AKMSA{sl}{2}{1}$ at levels~$1$ and~$-1/2$.
Characters of all these extensions turned out to be mock Jacobi forms for atypical modules~\cite{AC}.
It is probable that the Takif\/f superalgebra of  $\AKMSA{sl}{2}{1}$ appears in extensions of the $\tgl$ CFT, and interesting mock modular-like objects are expected to appear as characters of extended algebra modules.

It is also obvious that more realistic applications of Takif\/f af\/f\/ine algebras and superalgebras will require detailed understanding of Takif\/f af\/f\/ine $\widetilde{\widehat{\mathfrak{sl}}(2)}$. It seems natural to start its investigation from harmonic analysis, as a mini superspace toy model. One can expect here more involved structures of modules under the action of left- and right-invariant vector f\/ields on the group manifold, since the representation theory of $\widetilde{\mathfrak{sl}(2)}$ consists of inf\/inite-dimensional modules. It is also natural to expect that its free f\/ield realization will contain usual Wakimoto $\widehat{\mathfrak{sl}}(2)$ realization embedded into it.

\subsection*{Acknowledgements}

A.~Babichenko is thankful for hospitality to DESY Hamburg Theory Group and to ETH Zurich Institute for Theoretical Physics, where the f\/inal part of this work was done.
His work was supported by SFB676, and his visit to~ETH~-- by Pauli Center for Theoretical Studies.
T.~Creutzig is supported by NSERC Research Grant (Project $\#$: RES0020460).

\pdfbookmark[1]{References}{ref}
\LastPageEnding

\end{document}